\documentclass[sigplan,nonacm]{acmart}
\makeatletter
\def\@shortauthors{Wang et al.}
\makeatother

\setcopyright{none}
\renewcommand\footnotetextcopyrightpermission[1]{}
\usepackage{tikz}
\usepackage{amsmath}

\usepackage[OT1]{fontenc}
\usepackage{algorithm}
\usepackage{algorithmicx}
\usepackage[noend]{algpseudocode}
\usepackage{textcomp}
\usepackage{booktabs} 
\usepackage{comment}
\usepackage{wrapfig}
\usepackage{xcolor}
\usepackage{graphicx}
\usepackage{array}
\usepackage{multicol}
\usepackage{afterpage}
\usepackage{float}
\usepackage{listings}
\usepackage{multirow}
\usepackage{arydshln}
\usepackage{xspace}
\usepackage{enumitem}
\usepackage{fancyhdr}
\usepackage{amsmath}
\usepackage{url}
\usepackage{todonotes}
\usepackage{dirtree}
\definecolor{FadedBanana}{RGB}{255,255,191}
\usepackage{makecell}

\usepackage[utf8]{inputenc}
\usepackage{newunicodechar}
\usepackage{lineno}

\usepackage{xurl}

\usepackage{titlesec}
\titlespacing*{\section}{0pt}{6pt plus 4pt minus 2pt}{4pt plus 2pt minus 2pt}
\titlespacing*{\subsection}{0pt}{6pt plus 3pt minus 2pt}{4pt plus 2pt minus 2pt}
\titlespacing*{\subsubsection}{0pt}{6pt plus 3pt minus 2pt}{4pt plus 2pt minus 2pt}
\usepackage{tikz}

\usepackage{cleveref}
\crefformat{chapter}{\S#2#1#3} 
\crefformat{section}{\S#2#1#3} 
\crefformat{subsection}{\S#2#1#3}
\crefformat{subsubsection}{\S#2#1#3}

\usepackage{subcaption}

\graphicspath{{figures/}}
\usepackage{pifont}

\newcommand{\phb}[1]{\noindent\textbf{#1}\hspace{.5em}} 
\newcommand{\phm}[1]{\vspace{.4em} \noindent\textbf{#1}\hspace{.5em}} 
\newcommand{\athena}{\textit{IGTCache}}

\title{
        Efficient Unified Caching for Accelerating Heterogeneous AI Workloads
	}

\author{%
  Tianze Wang\textsuperscript{1},\quad
  Yifei Liu\textsuperscript{1},\quad
  Chen Chen\textsuperscript{1},\quad
  Pengfei Zuo\textsuperscript{2},\quad
  Jiawei Zhang\textsuperscript{2},\\
  Qizhen Weng\textsuperscript{3},\quad
  Yin Chen\textsuperscript{3},\quad
  Zhenhua Han\textsuperscript{4},\quad
  Jieru Zhao\textsuperscript{1},\quad
  Quan Chen\textsuperscript{1},\quad
  Minyi Guo\textsuperscript{1}\\[1ex]
  \textsuperscript{1}\,\textit{Shanghai Jiao Tong University}\quad\quad
  \textsuperscript{2}\,\textit{Huawei Cloud} \\
  \textsuperscript{3}\,\textit{Institute of Artificial Intelligence (TeleAI), China Telecom}\quad\quad
  \textsuperscript{4}\,\textit{Unaffiliated}
}

\begin{abstract}

Modern AI clusters, which host diverse workloads like data pre-processing, training and inference, often store the large-volume data in cloud storage and employ caching frameworks to facilitate remote data access.
To avoid code-intrusion complexity and minimize cache space wastage, it is desirable to maintain a unified cache shared by all the workloads. 
However, existing cache management strategies, designed for specific workloads, struggle to handle the heterogeneous AI workloads in a cluster---which usually exhibit heterogeneous access patterns and item storage granularities.
In this paper, we propose \textit{IGTCache}, a unified, high-efficacy cache for modern AI clusters.
\textit{IGTCache} leverages a hierarchical access abstraction, \texttt{AccessStreamTree}, to organize the recent data accesses in a tree structure, facilitating access pattern detection at various granularities. 
Using this abstraction, \textit{IGTCache} applies hypothesis testing to categorize data access patterns as sequential, random, or skewed. Based on these detected access patterns and granularities, \textit{IGTCache} tailors optimal cache management strategies including prefetching, eviction, and space allocation accordingly. 
Experimental results show that \textit{IGTCache} increases the cache hit ratio by 55.6\% over state-of-the-art caching frameworks, reducing the overall job completion time by 52.2\%.

\end{abstract}

\begin{document}

\maketitle

\section{Introduction}
\label{sec:intro}

AI innovations demand diverse computing workloads like data pre-processing~\cite{xu2021dp,salau2019feature}, model training~\cite{ray2019quick,dong2021survey}, and serving~\cite{zhou2024survey,yu2022survey}, which are usually hosted in cutting-edge computing clusters and rely heavily on various forms of data. 
Since the volumes of typical AI datasets are persistently increasing, for cost-efficiency and space elasticity, it is a common practice to store them in cloud storage like AWS S3~\cite{S3}---separated from computing infrastructures~\cite{krichevsky2021quantifying,jalaparti2018netco}. 
To mitigate the overhead of accessing remote cloud storage, caching frameworks like Alluxio~\cite{Alluxio} and JuiceFS~\cite{JuiceFS} are widely adopted. 
For high caching efficacy (i.e., high cache hit ratio and fast execution of the upper-level AI workloads), cache management policies---including prefetching, eviction and allocation---are of paramount significance. 

Specifically, while many optimized cache systems have been developed for specific workloads like model training or big data analytics, it is however not a good choice to associate each AI workload with a dedicated cache. 
First, having each workload maintain its own cache system introduces a significant code-intrusion burden. 
Meanwhile, given that AI datasets may be shared by different workloads, maintaining a separated cache for each workload would cause caching redundancy. 
Moreover, reserving cache space exclusively for each workload also suffers the internal fragmentation problem.  
Therefore, for AI clusters, we need a unified cache that can serve diverse workloads in a pluggable manner. 


However, it is challenging to attain high caching efficacy with such a unified cache.
Existing cache policy innovations primarily focus on specific data access scenarios.
For example, for prefetching, some works~\cite{al2020effectively,gill2007amp} focus on block-level prefetching by detecting the sequential access pattern, and some others~\cite{kroeger2001design,al2015predictive} focus on file-level prefetching by analyzing the historical request correlations.
For eviction, the LRU policy~\cite{o1993lru,smaragdakis1999eelru} and its variants~\cite{qureshi2007adaptive,megiddo2003arc} are commonly adopted for conventional workloads like big data, yet for model training workloads, recent works propose a better uniform-caching policy~\cite{mohan2020analyzing,zhao2023silod}.
These methods are tailored towards specific access patterns (e.g., sequential or random) and data storage granularities (e.g., file or block), not appropriate for the unified cache in AI clusters.

Specifically, with diverse AI workloads concurrently served by a cluster, the unified cache must simultaneously deal with heterogeneous data access patterns and granularities.
On the one hand, the workloads for AI development, including dataset-processing, training and inference, exhibit heterogeneous data access patterns: sequential for inferences, random for training, and skewed otherwise for workloads like data augmentation~\cite{deng2024lakebench} or RAG queries~\cite{lewis2020retrieval}. 
On the other hand, the data items of these workloads are often stored in different granularities, such as large text files (for corpus datasets)~\cite{zhu2015aligning,rajpurkar2016squad}, individual small files (for image, speech, and video datasets)~\cite{everingham2010pascal,Voxforge.org}, and multiple directories~\cite{deng2009imagenet,cisl_rda_dsd548000}.
Such heterogeneities in data access patterns and storage granularities render existing caching strategies not generally effective.
For example, our empirical studies in \Cref{sec:prior} show that, adopting block-level prefetching misses a 78.3\% speedup opportunity for a ResNet testing workload, yet turning to file-level prefetching misses a 22.7\% speedup for a BookCorpus pre-processing workload; besides, the cache hit ratio under LRU is 27.5\% higher than uniform-caching for a RAG workload, but it is 34.8\% lower when applied for a ResNet training workload.

Therefore, to realize a unified, high-efficacy cache for AI clusters, we need to 
equip it with the capability to adapt to the runtime data access patterns and granularities of each workload. 
We note that it is indeed feasible given three opportunities.
First, the cache frameworks can observe all remote data requests, supporting in-depth pattern analysis.
Second, data accesses in AI workflows exhibit strong locality, meaning that there usually exists a stable 
pattern among the data accesses to a data path. 
Third, the typical data access patterns are indeed rather limited (sequential, random and skewed), and it is possible to customize the caching policies respectively for each pattern.

To address the problem, this paper proposes \athena, a unified, high-efficacy cache for clusters containing heterogeneous AI workloads. 
\athena\ works with three key techniques. 
First, data access sequences from different workloads often mix up, making it hard to distinguish the correlated accesses from others; \athena\ thus introduces an abstraction named \texttt{AccessStreamTree} to organize the recent access history into a hierarchical structure, which facilitates access pattern detection at any possible granularity. 
Each node (\texttt{AccessStream}) in \texttt{AccessStreamTree} represents an abstracted data access stream at a specific granularity, and is treated as an independent unit for cache policy optimizations. 
Second, when analyzing the data access pattern for each \texttt{AccessStream}, handcrafted heuristics may fail to distinguish workloads with random or skewed patterns---which exhibit similar behavior when observed in a short window;
to address that problem, we apply hypothesis testing~\cite{an1933sulla,smirnov1948table} in access pattern recognition, which is accurate and statistically robust. 
Finally, based on the identified access pattern, \athena\ customize the cache management policies (e.g., prefetching, eviction, and allocation) respectively for each \texttt{AccessStream}.

We have implemented \athena\ atop JuiceFS~\cite{JuiceFS}, a popular file system for the cloud, with over 5,000 lines of Go code. 
Given a mixed workload suites including data-processing, training and inferences, our experimental results show that, \athena\ improves the cache hit ratio by 55.6\% over mainstream caching frameworks and reduces the average job completion time by 52.2\%.
In particular, each of the core functionalities, i.e., prefetching, eviction, and allocation, demonstrates substantial performance improvement from \athena. 
Meanwhile, our measurements show that the overhead of \athena\ is indeed negligible, with its computation accounting for only 0.36\% of the average I/O time. 

In summary, this paper makes the following contributions:
\begin{itemize}
    \item	
    We identify the need to design a unified cache for modern AI clusters, for which existing cache management practices, which are respectively designed for specific workload pattern, fail to attain high efficacy. 
    \item We propose \athena, a unified cache that organizes data accesses in an \texttt{AccessStreamTree}, makes accurate access pattern recognition for each \texttt{AccessStream} and adapts the caching policy for the best efficacy. 
    \item We implement \athena\ atop JuiceFS, and confirm its effectiveness with testbed experiments, demonstrating a 52.2\% improvement over existing caching practices across diverse AI workloads.
\end{itemize}

\section{Background and Motivation}
\label{sec:background}

\subsection{Cache: A Crucial Interlayer in AI Clusters}
\label{sec:trends}


\phb{Disaggregated storage architecture in AI clusters.}
AI technologies have substantially revolutionized various fields in the modern society~\cite{perrault2024artificial,toner2024artificial,rashid2024ai}.
For the development of new AI techniques, diverse AI workloads---like dataset pre-processing~\cite{xu2021dp,salau2019feature}, model training~\cite{ray2019quick,dong2021survey} and inference~\cite{zhou2024survey,yu2022survey}---are frequently launched in powerful computing clusters (we call AI clusters). 
Meanwhile, the data volume demanded by such AI workloads (e.g., to store the datasets or the model checkpoints) are increasingly expanding.
For example, GPT-4 is trained on a PB-level multi-modal dataset comprising text, images, and video~\cite{gpt4o,gpt4_size}, and the recently developed DeepSeek-V3 model requires 404 GB to store its parameters~\cite{deepseekollama,liu2024deepseek}. 

Consequently, it is now increasingly common to store the data on cloud storage platforms like S3~\cite{S3} and Azure Blob~\cite{Blob}, which are cost-effective and also scalable~\cite{elasticity-of-object-storage}. 
Such an architecture is called \textit{compute-storage disaggregation}, meaning that organizations maintain  compute resources locally while relying on remote storage for data access~\cite{kumar2020quiver, krichevsky2021quantifying}.
A recent study~\cite{satija2025cloudscape} on Amazon AWS shows that, among the machine learning (ML) services on the SageMaker platform (a general purpose model training and inference platform), over 58\% use S3 for reading training data and storing models.

\begin{figure}
    \centering
    \includegraphics[width=0.35\textwidth]{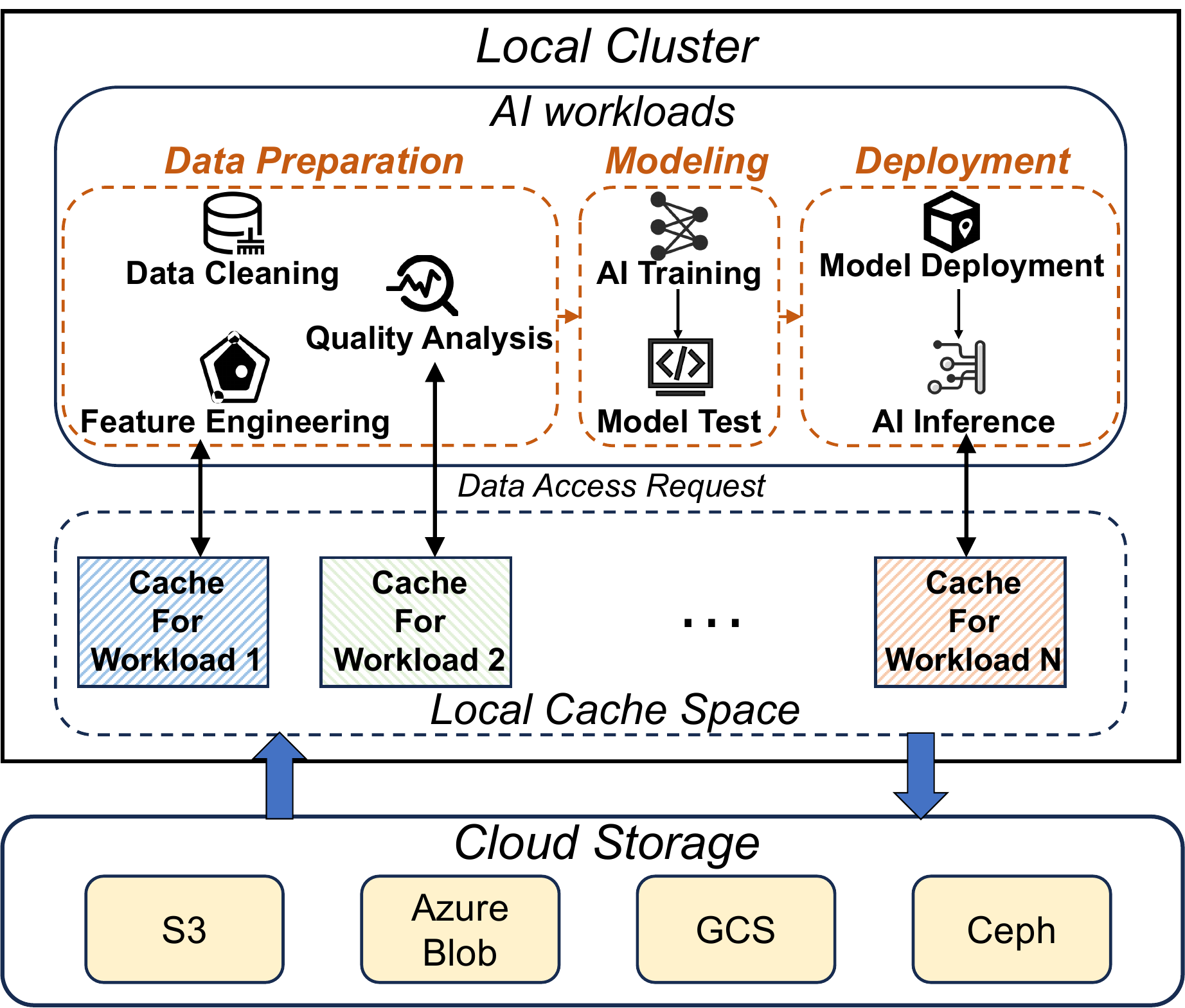}
    \caption{Workloads, cache and storage in AI clusters.}
    \label{fig:func}
\end{figure}

\phm{Caching frameworks to handle disaggregated storage.}
Accessing cloud storage like S3, however, incurs significantly higher data access latency~\cite{gao2016network,yelam2022systems} and bandwidth consumption.
To mitigate such overhead, production caching frameworks like Alluxio~\cite{Alluxio} and JuiceFS~\cite{JuiceFS} are increasingly adopted as an intermediate layer between AI computations and remote data store. 
As shown in \Cref{fig:func}, these frameworks cache frequently accessed data locally within the cluster, enabling direct access without remote communication.


Compared with the huge data volume on remote storage, the caching space in AI clusters (e.g., the memory space of the cache servers) is usually quite limited, thus requiring careful management to ensure efficient execution of the upper-level AI workloads. 
Specifically, cache management typically involves three aspects: prefetching, eviction, and space allocation.
Prefetching means to fetch in advance the data items that are predicted to be accessed in the future~\cite{al2020effectively}.
Eviction chooses the data items to kick out when there is no idle space~\cite{mohan2020analyzing}.
Space allocation means to set the amount of cache space allowed to be used by each workload
\cite{kumar2020quiver,gu2022fluid,gu2023adaptive}.
Those policies are crucial to the overall caching performance of the AI clusters. 

\phm{The need for a unified, high-performance cache for AI clusters.}
For high resource utilization, AI clusters are typically shared by diverse AI workloads.
As shown in \Cref{fig:func}, developing a new model typically requires three key stages: data preparation, modeling, and deployment, each including multiple workload types.
While a series of domain-specific cache optimization methods have been proposed in the literature~\cite{al2015predictive,mohan2020analyzing,khan2023shade,dryden2021clairvoyant}, we note that associating each workload with an isolated cache is suboptimal for several reasons.

First, requiring each workload to independently maintain its cache introduces a significant code-intrusion burden.
For example, adopting Quiver~\cite{kumar2020quiver}, a caching system optimized for training workloads, requires re-writing the data-feeding modules. 
Second, maintaining an isolated cache space for each workload would cause space wastage, because different workloads may manipulate on the same dataset~\cite{mihailescu2013mixapart,kakaraparthy2019case,mohan2020analyzing,kumar2020quiver}.
An analysis of Microsoft’s training workload trace~\cite{kakaraparthy2019case} shows that, due to the prevalence of dataset sharing, unifying data loading can reduce up to 89\% I/O overheads. 
Moreover, if launching an isolated cache for each workload (in popular frameworks like Alluxio~\cite{Alluxio}, the allocated cache space is exclusively reserved), the idled caching space of one workload is not reusable by others, incurring internal fragmentation. 

Therefore, in this paper we focus on designing a unified, high-performance cache shared by all the workloads within an AI cluster.
Such a unified cache shall work as a foundation caching service without requiring any application code intrusions.
In fact, popular caching frameworks like Alluxio~\cite{Alluxio} and JuiceFS~\cite{JuiceFS} already offer a unified data request interface (FUSE), which allows using conventional POSIX APIs~\cite{juicefsfuse,alluxiofuse} to access remote data stored in diverse backends.
In that sense, a unified cluster cache is already feasible at the API level.
However, our subsequent analysis reveals that, when confronted with the diverse workloads in modern AI clusters, it remains challenging for a unified cache to achieve high caching efficacy.


\subsection{Limitations of Existing Caching Methods} 
\label{sec:prior}

Cache management is a classical problem in the literature (in domains like operating system, big data and machine learning), for which many prefetch, eviction and allocation methods have been proposed. 
Here we review those methods and discuss their limitations for our problem. 

\phm{Prefetching.} 
Data prefetching can be broadly classified into block-level prefetching and file-level prefetching.
At the block level, some works~\cite{al2020effectively,gill2007amp} conduct prefetching for sequential access patterns, and some others prefetch correlated blocks via history mining~\cite{yang2017mithril,li2004c} or leveraging system/application hints~\cite{soundararajan2008context}.
At the file level, prefetching is conducted mainly by analyzing the correlations among frequently accessed files~\cite{kroeger2001design,al2015predictive}.
Meanwhile, for model training workloads, caching frameworks like Alluxio~\cite{Alluxio} can preload the dataset into the cache space with an explicit user command~\cite{data_preloading_alluxio}, while some other works~\cite{dryden2021clairvoyant,khan2023shade} propose to use application-side information (e.g., the random seed) for prefetching.

\phm{Eviction.} Most eviction strategies adopted in practice are built upon classical principles such as LRU~\cite{o1993lru,smaragdakis1999eelru,qureshi2007adaptive}, LFU~\cite{einziger2022lightweight,einziger2017tinylfu}, and FIFO~\cite{yang2023fifo,zhang2024sieve}.
Besides, caching frameworks like Alluxio and JuiceFS also employ a \emph{time-to-live} (TTL) hyperparameter to enable proactive data removal from cache~\cite{alluxio_ttl,juicefs_ttl}, which is more efficient to implement than LRU. 
For model training workloads, the \emph{uniform-caching} strategy~\cite{mohan2020analyzing,zhao2023silod} is typically used, under which accessed data samples are pinned in the cache until capacity is reached and are not evicted thereafter.

\phm{Allocation.} Static cache allocation assigns cache space to each dataset based on application-level information, such as total batch size~\cite{gu2022fluid}. 
Dynamic cache allocation adapts by estimating the cache demand for different workloads at runtime.
Specifically, Cuki~\cite{gu2023adaptive} and SlidingSketch~\cite{gou2020sliding} focus on efficiently and accurately estimating the working set size for big data workloads.
For model training workloads, which often exhibit random access patterns during repetitive epochs, existing methods estimate cache demand with the dataset size and data consumption speed~\cite{kumar2020quiver,zhao2023silod}. 


		
To summarize, existing cache management strategies are designed for specific workload types or data storage granularities. 
Those methods, however, fail to make high caching performance when adopted by the unified cache of an AI cluster---due to its built-in heterogeneity in both data storage granularity and data access patterns.
Below we detail such heterogeneity and empirically verify the incapability of existing caching methods when confronting such heterogeneity. 

\begin{table}[]
\centering
\resizebox{\columnwidth}{!}{%
\begin{tabular}{ccc}
\toprule
Dataset & Storage Semantics & Storage Granularity \\ \midrule
BookCorpus~\cite{zhu2015aligning} & train/data-\{id\}.arrow & 74M records in 16 files \\ \hline
SQuAD~\cite{rajpurkar2016squad} & data/cached\_data.pth & 157K records in one file \\ \hline
PASCAL-VOC~\cite{everingham2010pascal} & JPEGImages/\{id\}.jpg & 17K images in one directory \\ \hline
VoxForge~\cite{Voxforge.org} & wav/\{user\}\_\{date\}\_\{id\}.wav & 95K audio files in one directory \\ \hline
ImageNet~\cite{deng2009imagenet} & \{class\}/\{id\}.jpg & Images in 1K category directories \\ \hline
ICOADS~\cite{cisl_rda_dsd548000} & \{date\}/\{coordinate\}.csv & Tables in 2K date-based directories \\ \hline
\multirow{2}{*}{COCO~\cite{lin2014microsoft}} & \multirow{2}{*}{\makecell{annotations/\{usage\}.json \\ train2017/\{id\}.jpg}} & Jsons in one directory  \\ 
 &  & 330K images in one directory \\ \hline
 \multirow{2}{*}{Flickr30k~\cite{plummer2015flickr30k}} & \multirow{2}{*}{\makecell{results\_20130124.token \\ flickr30k-images/\{id\}.jpg}} & 158K records in one file  \\
 &  &  30K images in one directory\\
\bottomrule
\end{tabular}%
}
\caption{Examples of datasets with different storage forms (COCO and Flickr30k are multi-modal datasets).}
\label{tab:different_storage_forms}
\end{table}

\phm{Heterogeneity of data storage granularities (\emph{block}, \emph{file} and \emph{directory}) and its impact on caching performance.} 
    AI innovations heavily rely on diverse datasets with massive data items stored in unstructured files. As shown in Table \ref{tab:different_storage_forms}, data items in different datasets may be organized in different granularities.
    In some cases (e.g., the BookCorpus~\cite{zhu2015aligning} and SQuAD~\cite{rajpurkar2016squad} datasets), the entire dataset is stored as a few large text files, in which each data item (e.g., a query-answer pair) spans less than one block. 
    In other cases (e.g., the PASCAL-VOC~\cite{everingham2010pascal} and VoxForge~\cite{Voxforge.org} datasets), the entire dataset is stored in a directory, with each data item (i.e., an image or a speech/video fragment) corresponding to a dedicated file. 
    Moreover, some datasets (e.g., ImageNet~\cite{deng2009imagenet} and ICOADS~\cite{cisl_rda_dsd548000} datasets) are organized in an array of directories, in which each directory contains a subset of data items based on category or source. 
    Therefore, a unified, high-performance cache must properly deal with such heterogeneous storage granularities. 

\begin{figure}[t]
\centering
    \subcaptionbox{Prefetching\label{deficiency:prefetch}}{\includegraphics[width = 0.22\textwidth]{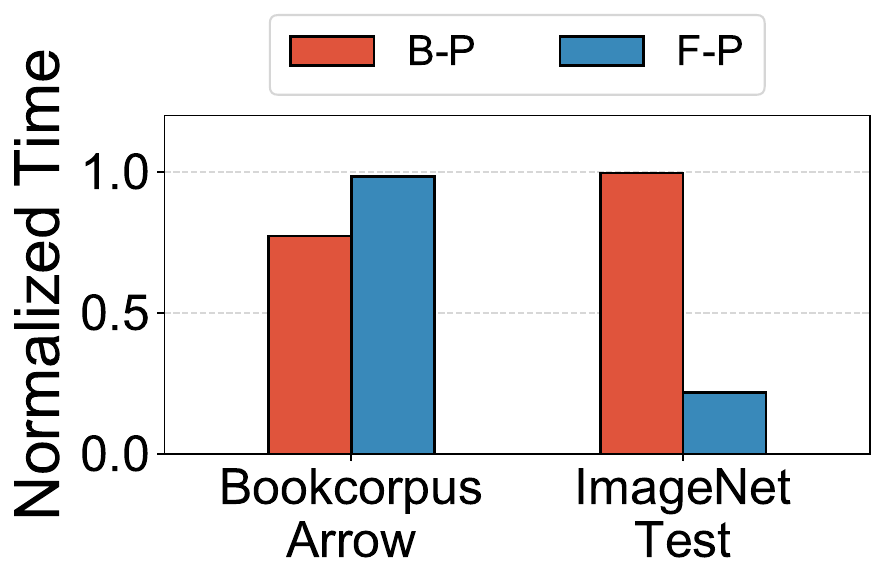}}
	\hspace{1mm}
	\subcaptionbox{Eviction\label{deficiency:eviction}}{\includegraphics[width = 0.22\textwidth]{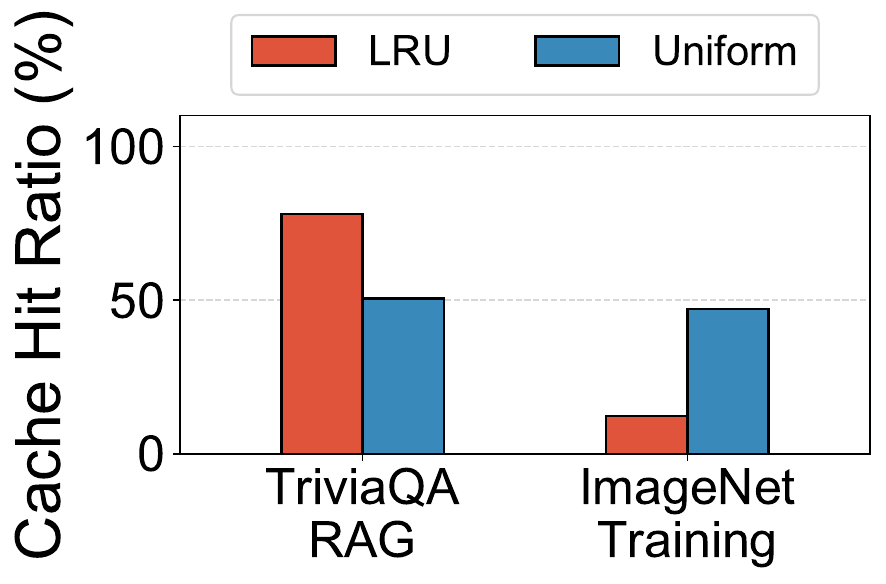}}
        \caption{
        (a) Block-level prefetching (B-P) speeds up pre-processing on Bookcorpus but fails to benefit ImageNet testing, whereas file-level prefetching (F-P) has the opposite effect. The times are normalized relative to the completion time without prefetching. (b) LRU eviction outperforms uniform-caching policy for the RAG workload but performs worse for ImageNet training.} 
\label{deficiency of inapproriate strategies}
\end{figure}

To evaluate the impact of granularity heterogeneity in caching (prefetching for example) performance, we store two datasets, BookCorpus~\cite{zhu2015aligning} and ImageNet~\cite{deng2009imagenet}, in S3. 
The BookCorpus dataset is a single file containing many data items each smaller than a block; yet the ImageNet dataset is composed of massive small files. 
Those datasets are accessed via the JuiceFS~\cite{JuiceFS} cache framework---a shared production caching framework with block-level prefetching by default.
We run text preprocessing workload on BookCorpus and testing workload on ImageNet, both on a V100 GPU in our local cluster. 
In \Cref{deficiency:prefetch}, we show the relative inference completion time normalized by that with no cache.
As shown in \Cref{deficiency:prefetch}, the default block-level prefetching speeds up the preprocessing over BookCorpus by 22.7\%, but has no benefit for ImageNet.
We then customize JuiceFS by applying prefetching at the file granularity. This adjustment speeds up the ResNet-50 test by 78.3\%, but the speedup for BookCorpus diminishes to 0.
This experiment highlights that prefetching methods designed for a specific granularity may not perform generally well for diverse datasets in AI clusters.

    

\phm{Heterogeneity of data access patterns (\emph{sequential}, \emph{random}, and \emph{skewed}) and its impact on caching performance.}
    As shown in \Cref{fig:func}, AI workloads involve multiple stages.
    From the cache point of view, these workloads exhibit diverse data access patterns, broadly classified as \emph{sequential}, \emph{random}, and \emph{skewed}.
        1) Sequential access patterns are common in preprocessing and inference tasks, where data samples are processed in a sorted manner. 
        2) Random access patterns are prevalent during model training, where data samples in each mini-batch are selected based on a random sequence generator. 
        3) Skewed access patterns mean that some items are accessed more frequently than others. 
        For example, AI query workloads such as data enrichment~\cite{deng2024lakebench} and retrieval-augmented generation\footnote{
        While a small-scale knowledge set may be directly stored within the GPU server, large-scale knowledge sets as retrieval sources may however be hosted on the cloud. Milvus~\cite{Milvus}, a powerful vector database for RAG, reports using S3~\cite{S3} to store the large files like index files and binary logs~\cite{milvuss3}.
        } (RAG)~\cite{lewis2020retrieval} often exhibit skewed patterns.
        Notably, the three access patterns are quite common for AI workloads and shall thus be treated all as first-class citizens in cache design. 

We further evaluate the impact of pattern heterogeneity on caching (eviction for example) performance.
We first run an RAG query workload using the TriviaQA~\cite{joshi2017triviaqa} dataset and then conduct ResNet-50 training on ImageNet, both via JuiceFS with LRU policy for eviction.
We set the cache size to half of the size of each dataset.
For comparison across different eviction methods, we further replace the default LRU policy with uniform caching.~\Cref{deficiency:eviction} shows the respective cache hit ratio in both cases. 
For the RAG workload, the cache hit ratio for LRU is 27.5\% higher than uniform caching. However, for ResNet-50 training, the cache hit ratio for LRU is 34.8\% lower.
These results suggest that eviction strategies need to account for data access patterns to achieve optimal cache performance.
Note that the mixture of different data access patterns also poses challenges to cache allocation.
Currently, cache allocation is conducted only among homogeneous workload types (e.g., purely among query workloads~\cite{gu2023adaptive} or purely among training workloads~\cite{kumar2020quiver}), and it remains unclear how to make cache allocation among heterogeneous workloads.

In summary, although existing caching strategies perform well under specific granularity-pattern combinations, they do not generalize well across the diverse storage granularities and access patterns in an AI cluster, as illustrated in \Cref{tab:prior_works}.
In that sense, to attain high efficacy, a unified cache for AI clusters must be able to automatically adapt its caching policy to the runtime data access pattern and data storage granularity.



\begin{table}
    \centering
    \Large
    \resizebox{\columnwidth}{!}{
        \begin{tabular}{|>{\centering\arraybackslash}p{2.0cm}|c|>{\centering\arraybackslash}p{1.0cm}>{\centering\arraybackslash}p{0.8cm}>{\centering\arraybackslash}p{1.0cm}|>{\centering\arraybackslash}p{1.2cm}>{\centering\arraybackslash}p{1.2cm}>{\centering\arraybackslash}p{1.2cm}|}
            \hline
            \multirow{2}{*}{Function} & \multirow{2}{*}{Solutions} & \multicolumn{3}{c|}{Granularities} & \multicolumn{3}{c|}{Patterns} \\ \cline{3-8}
            &                                                                                                                           & \multicolumn{1}{c}{Block} & \multicolumn{1}{c}{File} & \multicolumn{1}{c|}{Dir.} & \multicolumn{1}{c}{Seq.} & \multicolumn{1}{c}{Rand.} & \multicolumn{1}{c|}{Skewed} \\ \hline
            \multirow{4}[0]{*}{Prefetch}   & ~\cite{Alluxio,JuiceFS,al2020effectively,garg2024crossprefetch} & \ding{51}& \ding{55}& \ding{55}& \ding{51}& \ding{55}& \ding{55}  \\ \cline{2-8}
            & ~\cite{yang2017mithril}                                                                                          & \ding{51}                 & \ding{55}                & \ding{55}                   & \ding{55}                      & \ding{55}& \ding{51}  \\ \cline{2-8}
            & ~\cite{dryden2021clairvoyant,khan2023shade}                                                    & \ding{55}                 & \ding{51}& \ding{55}& \ding{55}& \ding{51}& \ding{55} \\ \cline{2-8}
            & ~\cite{al2015predictive,wang2021sfp}                                                                    & \ding{55}                 & \ding{51}                & \ding{55}& \ding{55}& \ding{55}& \ding{51}\\ \hline

            \multirow{3}[0]{*}{Eviction}   & ~\cite{o1993lru,qureshi2007adaptive,yang2023fifo,zhang2024sieve} & \ding{51}& \ding{55}& \ding{55}& \ding{55}& \ding{55}& \ding{51}  \\ \cline{2-8}

            & ~\cite{mohan2020analyzing}                                                                                  & \ding{55}                 & \ding{51}                & \ding{55}                   & \ding{55}& \ding{51}& \ding{55}  \\ \cline{2-8}
            & ~\cite{alluxio_ttl}                                                                                       & \ding{55}                 & \ding{51}                & \ding{51}                   & \ding{55}                      & \ding{55}& \ding{51} \\ \hline

            \multirow{2}[0]{*}{Allocation} & ~\cite{mvondo2021ofc,romero2021faa,gu2023adaptive} & \ding{51} & \ding{51}& \ding{51}& \ding{55}& \ding{55}& \ding{51}  \\ \cline{2-8}
            & ~\cite{kumar2020quiver,gu2022fluid}                                                                   & \ding{51}                 & \ding{51}                & \ding{51}& \ding{55}& \ding{51}& \ding{55}  \\ \hline
            All & \athena & \ding{51} & \ding{51} & \ding{51} & \ding{51} & \ding{51} & \ding{51} \\ \cline{1-8}
        \end{tabular}%
    }
    
    \caption{Performance of existing caching strategies and ours across different storage granularities and access patterns (\ding{51}~indicates good performance and \ding{55}~indicates poor).}
    \label{tab:prior_works}
    
\end{table}%

\subsection{Opportunities and Challenges}
\label{subsec:challenges}

\phm{Opportunities.} 
It holds great promise to design a unified and high-performance cache for AI clusters---following an observe-and-act online adapting manner.

First, caching frameworks have the ability to observe historical data access information.
This comprehensive observation enables in-depth analysis on the runtime access patterns. 

Second, the massive unstructured data aligned well within a dataset, just like the memory address, often exhibit temporal and spatial locality\footnote{
\emph{Temporal} and \emph{spatial} data locality have long been adopted for cache optimizations~\cite{gu2022fluid,yang2017mithril,gu2023adaptive}.
Temporal locality means that, after a data item is accessed, the same item would probably be accessed again in the near future; spatial locality means that, the data items in nearby space of a just-accessed item would probably also be accessed in the near future.
}, because all the data items in an AI dataset are usually processed under the same programming logic (e.g., training or testing).
With a small observation window at execution commencement, we can discern the common access pattern applicable to the entire dataset.

Third, although there are diverse workloads in shared AI clusters, they can be well captured by three access patterns: sequential,
random and skewed.
Given such limited patterns, we can prepare solution kit respectively for each pattern, and select the solution kit for each runtime workload based on the recognized pattern. 


\phm{Challenges.}
While promising, realizing automatic cache management in practice is nonetheless a non-trivial task, and the challenges are as follows: 

    \textit{Challenge 1:} A production cache often concurrently serves requests from multiple AI workloads. 
    While the data access pattern of an AI workload is relatively clear, multiple access sequences from different workloads may mix up with each other, making it hard to distinguish the correlated accesses from others. 
    Meanwhile, a meaningful access pattern (e.g., sequential or random) may only appear at a certain granularity (e.g., file or block). For generality, we need to properly organize the historical data accesses to allow for efficient pattern recognition at any arbitrary granularity.
    
     \textit{Challenge 2:} Making prompt and also accurate pattern recognition is difficult in practice. From the cache point of view, the data access list of different patterns (i.e., random and skewed) may not be remarkably different (\Cref{sec:diagnosis}), and 
     naive handcrafted pattern recognition heuristics may fail to work consistently well in practice. 
    
     \textit{Challenge 3:} Even if the cache access pattern is accurately acquired, it is still challenging to adapt the caching policy for the best caching performance. For example, in some circumstances, naively applying sequential prefetching at directory-granularity would risk prefetching many unnecessary files (see later in \Cref{fig:hierarchical_prefetch}).
    Besides, pattern awareness itself does not answer how to optimize the cache efficacy by conducting inter-workload cache space migration.
    We need to systematically explore the potential design space so as to maximize the performance benefit of adaptive caching. 

We will address those challenges in the next section. 

\section{The IGTCache Design}
\label{sec:solution}


\begin{figure}
    \centering
    \includegraphics[width=0.5\textwidth]{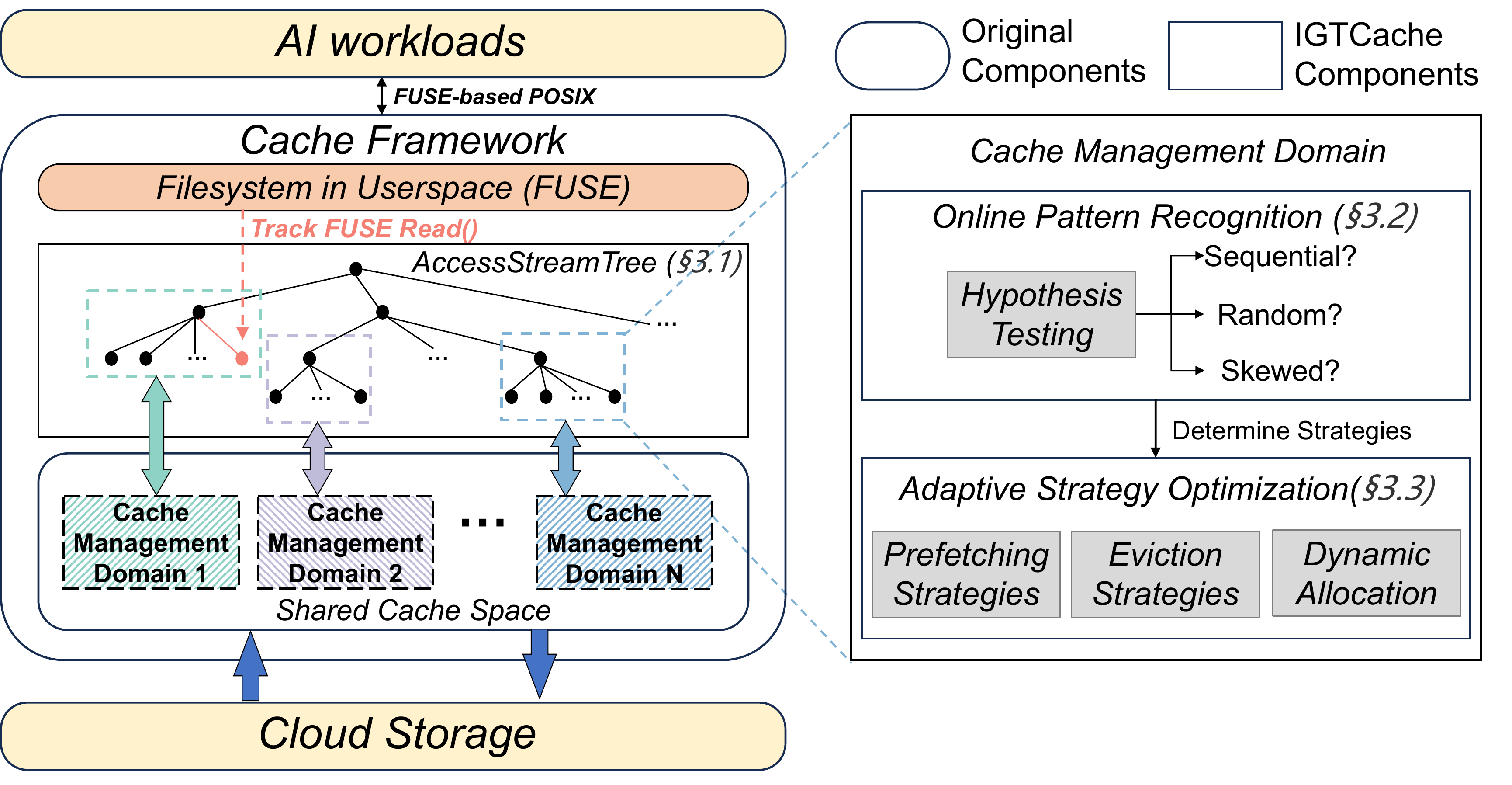}
    \caption{Architecture of \athena.}
    \label{architecture}
\end{figure}
In this section, we describe \athena, a unified, high-efficacy cache for heterogeneous AI workloads in modern AI clusters. 
As shown in \Cref{architecture}, the \athena\ design is composed of three parts.
First, to support pattern analysis at any potential storage granularity, \athena\ adopts an abstraction called \texttt{AccessStreamTree} to organize the historical accesses hierarchically, addressing \textit{Challenge 1} (\Cref{sec:AccessStreamTree}).
Second, for prompt and accurate pattern recognition, \athena\ introduces a hypothesis testing method to identify the access pattern for each workload at runtime, addressing \textit{Challenge 2}  (\Cref{sec:diagnosis}).
Third, for high efficacy, it customizes the prefetching, eviction, and space allocation strategies based on the workload types, addressing \textit{Challenge 3} (\Cref{sec:adaptive_management}).

\subsection{Hierarchical Access Abstraction}
\label{sec:AccessStreamTree}



To handle mixed AI workloads, it is essential to distinguish and organize correlated data accesses for ease of pattern recognition.
Due to the diverse storage granularities of different AI datasets, access patterns may appear at any granularity (directory, file or block, as elaborated in \Cref{sec:prior}).
To achieve this, we introduce the \texttt{AccessStreamTree} abstraction, which organizes recent data accesses into a tree\footnote{
    Some existing works~\cite{griffioen1994reducing,lei1997analytical,kroeger2001design} also proposed to organizing the historical data accesses in a tree structure. 
    Yet, our \texttt{AccessStreamTree} here has several key differences.
    First, the access trees in existing works usually involve only one specific storage granularity (e.g., either block or file for correlation analysis), yet our \texttt{AccessStreamTree} is more complete and can capture the access patterns at any granularity (directory, file and block). 
    Moreover, each node in our \texttt{AccessStreamTree} has integrated flexible functionalities. Each \texttt{AccessStream} independently analyzes its access patterns, customizes its caching policies and also works as a space isolation unit.   
} structure 
(for unified management, a single \texttt{AccessStreamTree} tracks accesses from all the workloads). 

\begin{figure}
    \centering
    \includegraphics[width=0.4\textwidth]{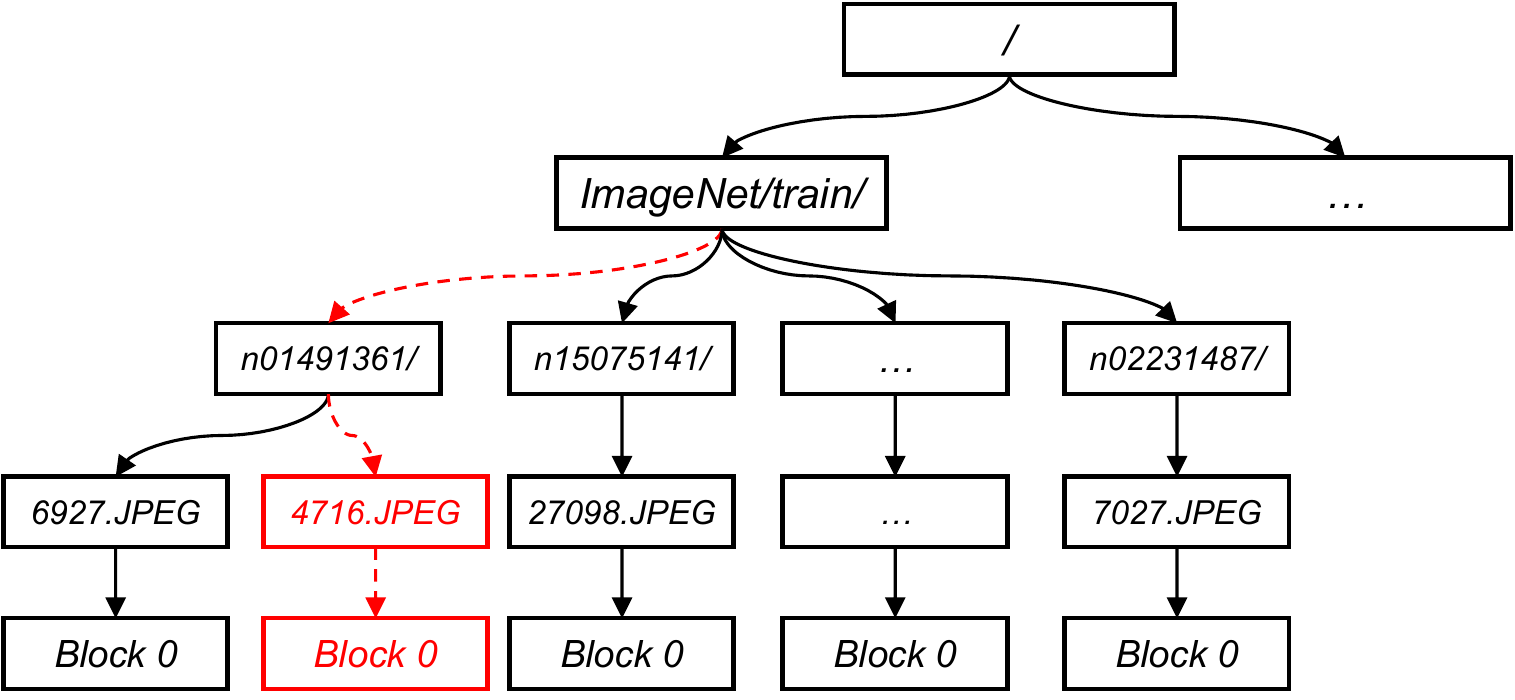}
    \caption{The structure of an \texttt{AccessStreamTree}.}
    \label{fig:AccessStreamTree}
    \vspace{-.1in}
\end{figure}

Specifically, each node in the \texttt{AccessStreamTree} represents an \texttt{AccessStream}, i.e., a conceptual unit grouping a set of accesses at a given level. 
Accesses within an \texttt{AccessStream} share the same path prefix and may have strong locality, and they form a unit for pattern analysis.
Moreover, an \texttt{AccessStream} is also a unit for cache policy customization: each \texttt{AccessStream} independently determines the prefetching candidates and the eviction policy for the data items it accessed, and it also works as a space isolation unit: when the total cache consumption of its data items is larger than allocated, it will conduct local replacement to comply with the  \texttt{AccessStream}-level allocation amount.

The workflow to build and utilize the \texttt{AccessStreamTree} is illustrated in \Cref{fig:AccessStreamTree}. 
For each block access (e.g., the first block for the file \texttt{/ImageNet/train/n01491361/4716.JPEG} in \Cref{fig:AccessStreamTree}), \athena\ gets its full path and adds it to the existing \texttt{AccessStreamTree} via prefix matching.
For each new access level along the path, a new \texttt{AccessStream} node is created (e.g., \texttt{4716.JPEG} and \texttt{Block 0} in \Cref{fig:AccessStreamTree}).
For each \texttt{AccessStream} node, once the number of its child nodes exceeds a predefined observation window size (defaulted to 100), it is classified as a \emph{non-trivial} node. At this point, \athena\ triggers pattern analysis (\Cref{sec:diagnosis}) and policy customization (\Cref{sec:adaptive_management}) at that very level.
Additionally, to control the overhead to maintain such an AccessStreamTree, we adopt a series of techniques (e.g., \emph{layer compression} and \emph{child pruning}), which will be elaborated later in \Cref{sec:implementation}.

\subsection{Online Pattern Recognition}
\label{sec:diagnosis}

\begin{figure}[t]
\centering
    \subcaptionbox{Random: ImageNet Train\label{access:random}}{\includegraphics[width = 0.22\textwidth]{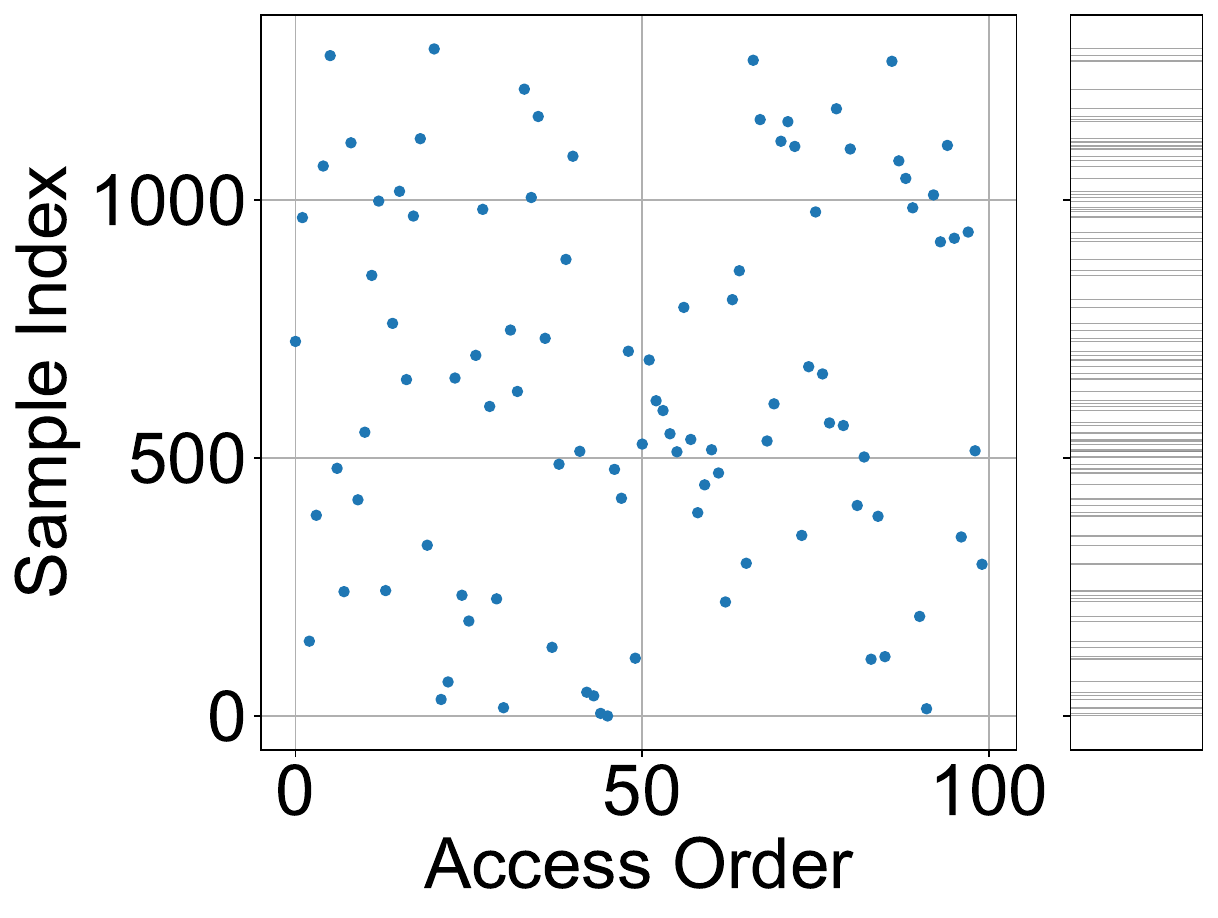}}
    \hspace{0.02\textwidth}
    \subcaptionbox{Skewed: TriviaQA~\cite{joshi2017triviaqa} RAG\label{access:skewed}}
    {\includegraphics[width = 0.22\textwidth]{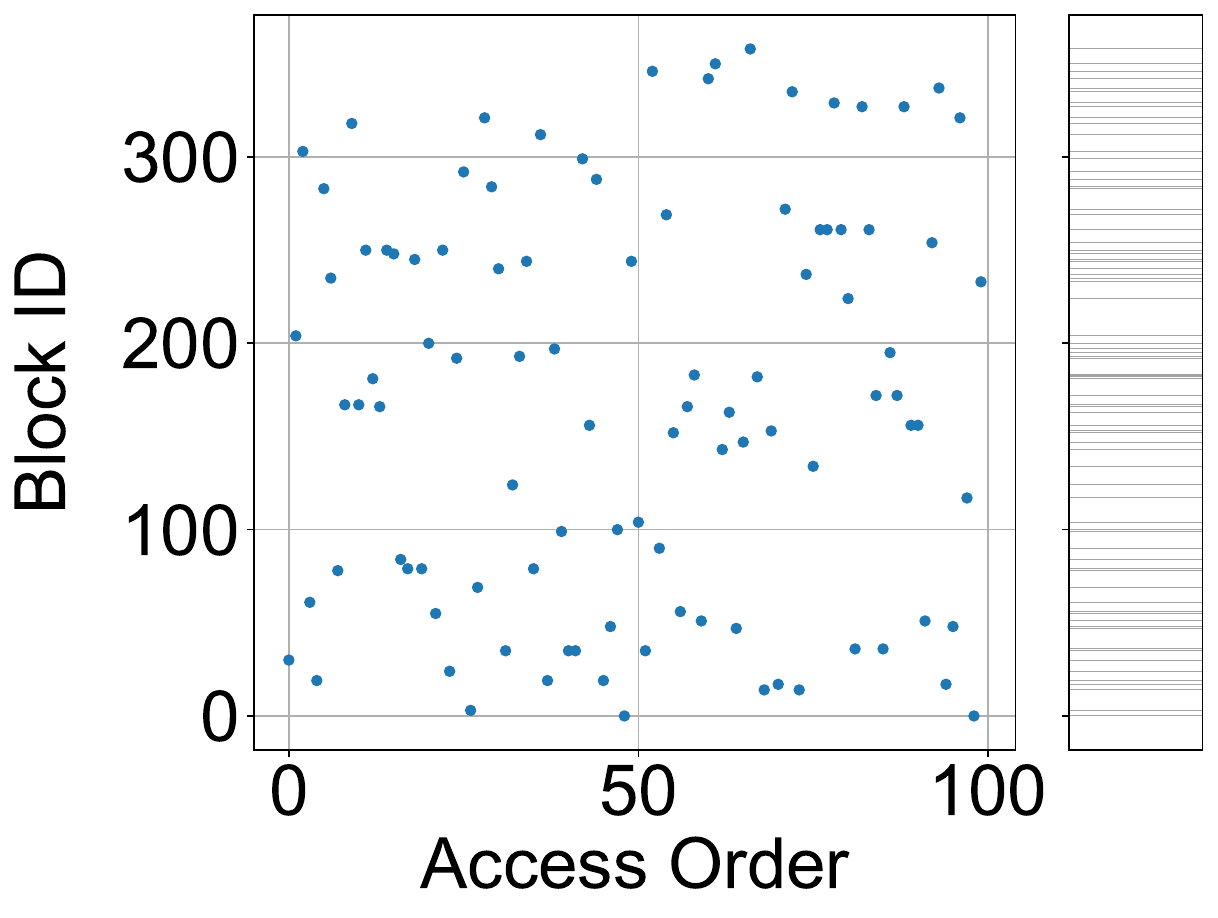}}
    \caption{Access sequences of random and skewed patterns. While the two patterns desire different caching policies, their access behaviors are however quite similar.} 
\label{fig:access features}
\end{figure}
For each non-trivial \texttt{AccessStream}, \athena\ maps it to one of the three access patterns: sequential, random, and skewed.
To avoid user-code intrusion, the pattern recognition must be made purely with the cache-side information, i.e., with the data access sequence in each \texttt{AccessStream}. 

We first note that it is straightforward to distinguish sequential patterns from others---by monitoring the spatial gaps of consecutive read requests, as in existing practice~\cite{wu2007linux,al2020effectively}. 
Specifically, we use the data item index to calculate the spatial gap; for a block that index is the block id, and for a file or directory, its index number can be obtained as the sequential element number\footnote{
In many datasets (as in \Cref{tab:different_storage_forms}), the file names themselves are rigidly formatted, often containing semantic information like data generating time or serial number. 
    In such cases, we can potentially also use string-processing functions to quantify the access gap with such semantic information.
} in the parent directory (i.e., the default access order at traversal time).
If an \texttt{AccessStream} is not sequential, we then proceed to check whether it follows the random or skewed pattern.

Yet, it is however not an easy task to distinguish between random and skewed patterns. 
To elaborate, as shown in \Cref{fig:access features}, while random and skewed access patterns desire fundamentally different caching policies (as recorded in \Cref{tab:prior_works}), their access behaviors during an observation window may be quite similar with each other.
In that sense, instead of designing ad-hoc heuristics with risky hyper-parameters and non-assured effectiveness, we seek to design a robust pattern recognition method with statistical confidence.

\begin{figure}[t]
\centering
    \subcaptionbox{ImageNet Training\label{gap:random}}{\includegraphics[width = 0.23\textwidth]{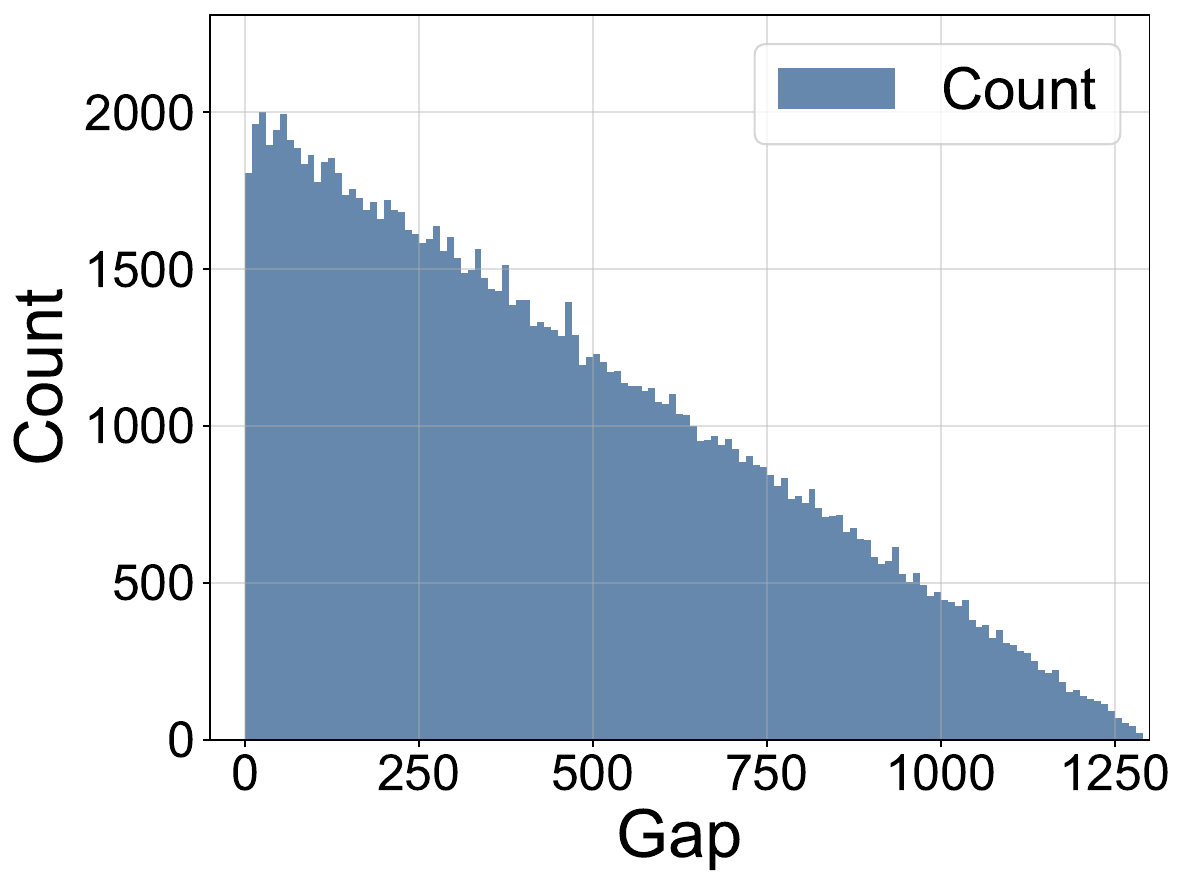}}
    \subcaptionbox{LakeBench Table-join\label{gap:table}}{\includegraphics[width = 0.23\textwidth]{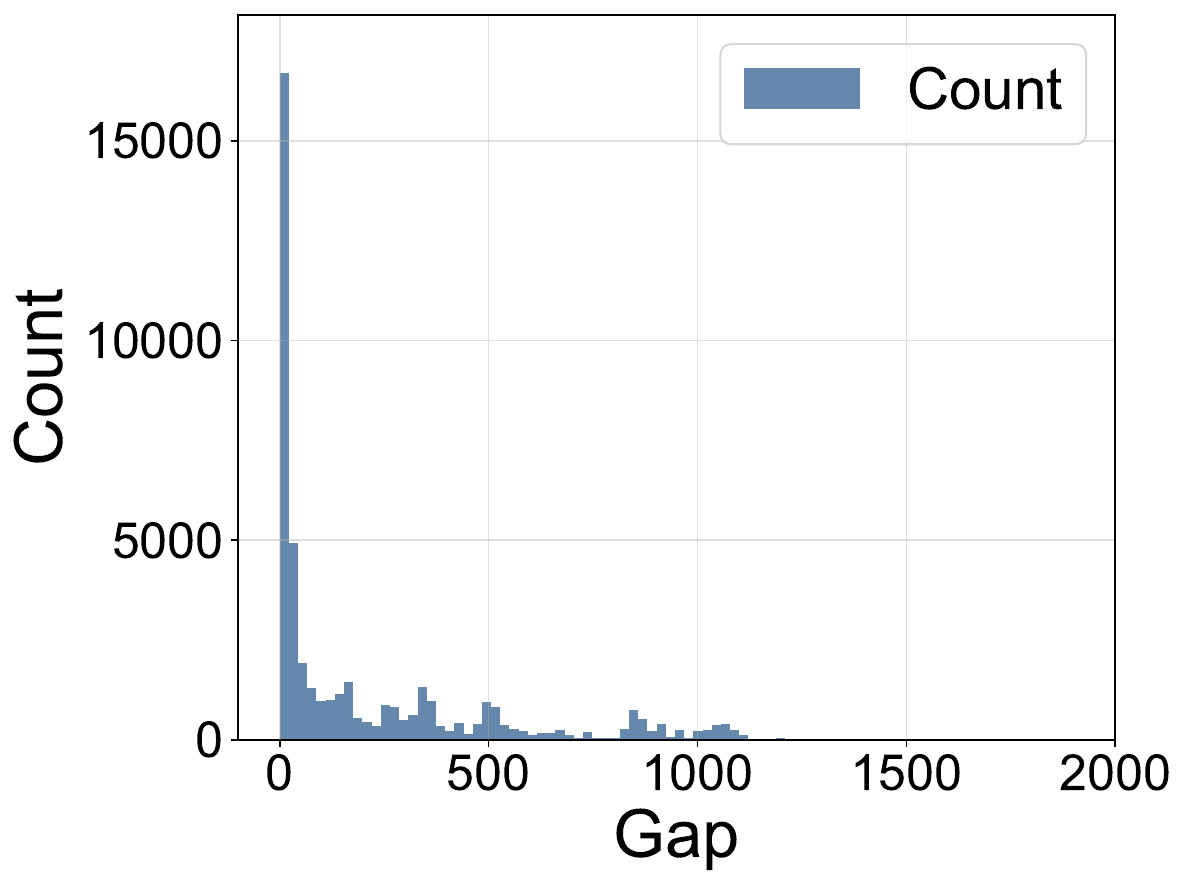}}
    \caption{Distributions of spatial gaps between any two consecutive accesses for workloads with different patterns. (a) Triangular shape for a training workload with random pattern. (b) Non-triangular shape for a query-based data augmentation workload~\cite{deng2024lakebench} with skewed pattern.}
\label{fig:spatial features}
\end{figure}

Given the locality among accesses of an \texttt{AccessStream}, we propose recognizing the access pattern based on the distribution of the spatial gaps between any two consecutive data accesses. 
In each non-trivial \texttt{AccessStream}, the spatial gap between consecutive accesses can be viewed as drawn from a fixed distribution.
By analyzing the shape of such distribution, we can infer the underlying access pattern.
To be specific, we let $Z$ be the random variable denoting the spatial gap between two consecutive accesses, 
then random and skewed patterns exhibit different $Z$ distribution shapes.

    
Specifically, for the random pattern, the index of each accessed data item can be viewed as sampled from a uniform distribution $[1,c]$, where $c$ is the total number of items in that dataset. 
    In this case, the spatial-gap ($Z$) distribution exhibits a \emph{triangular} shape, as shown in \Cref{gap:random} with a training workload. Mathematically\footnote{
        The derivation process is straightforward: given any two consecutive sample requests, there are $c(c-1)$ different index combinations (with repeating access excluded because in one epoch each sample is accessed only once) Among them, there are $2(c-k)$ combinations with a gap of $k$.
    }, the \emph{probability mass function} (PMF) of the distribution is $P(Z=k)=\frac{2(c-k)}{c(c-1)}$.
In contrast, the skewed pattern covers all the workloads that do not exhibit purely sequential or random patterns. Therefore, its spatial-gap distribution may exhibit any arbitrary shape other than {impulse} or {triangular}. 
    For example, for the LakeBench dataset, the distribution follows a long-tail shape as shown in~\Cref{gap:table}.

To summarize, recognizing the access pattern involves judging the shape of the entire $Z$ distribution with the available samples in the observation window. 
This is essentially a \emph{hypothesis-testing} problem, where the judgment is made via the \emph{test statistic} and the \emph{significance level}. 
In particular, our objective is to judge whether a collection of samples is drawn from a reference distribution, 
we choose the \emph{Kolmogorov–Smirnov test method} (K-S test)~\cite{an1933sulla,smirnov1948table} for hypothesis testing.
In the K-S test, the \emph{null hypothesis} is that the samples are drawn from the reference distribution.
The test then quantifies the maximum difference between the \emph{empirical cumulative distribution function} (ECDF) of the samples and the \emph{cumulative distribution function} (CDF) of the reference distribution, which is used as the test statistic $D_{max}$.
After that, given the significance level $\alpha$ (typically set to 0.01) and the number of samples, we can get the K-S test threshold $D_\alpha$ by searching a reference table.
If $D_{max}<D_\alpha$, we cannot refuse the null hypothesis at the significance level, meaning the pattern assumption holds.

Specifically, when checking whether the accesses follow a random pattern, since the PMF of that triangular distribution is $P(Z=k)=\frac{2(c-k)}{c(c-1)}$, we can derive that the CDF of the reference distribution is: 
\begin{equation}
F(k) = \sum_{j=1}^{k} P(Z = j) = \frac{2k}{c-1} - \frac{k(k+1)}{c(c-1)}, 1\leq k\leq c-1.
\label{eq:cdf_sum}
\end{equation}
We then use the K-S test to decide whether to accept the hypothesis.
If yes, we conclude that the accesses in the \texttt{AccessStream} follow a random pattern\footnote{
When multiple training workloads share the same datasets, the cumulative data access pattern in that \texttt{AccessStream} is still random. 
Meanwhile, we do not consider the complex case with mixed random and non-random patterns (e.g., if training and data pre-processing are simultaneously conducted on a dataset), which threatens data consistency and is rare in practice.
}. Otherwise, the accesses follow the skewed pattern.
Note that since the observation window is usually small (defaulted to 100), the runtime overhead of the K-S test is negligible, making it suitable for online use (which we will evaluate later in \Cref{sec:eval_overheads}). 

\subsection{Adaptive Cache Strategy Optimization}
\label{sec:adaptive_management}
In this subsection, we show how to optimize the cache management strategies based on both the storage granularities and recognized access patterns.
Our strategy optimizations involve prefetching, eviction, and space allocation. 

\begin{sloppypar}
\phm{Prefetching Strategy Optimization.}
For each \texttt{AccessStream}, we enhance its prefetching effectiveness with both pattern adaptivity and granularity adaptivity.
Regarding pattern adaptivity, \athena\ automatically switches the prefetching policy\footnote{Note that our focus here is to enable the adaptivity to switch between different policies instead of to enhance each individual policy; the policy suite here can be flexibly extended to cover more sophisticated ones.
} to the best one for each pattern. 
For the \emph{sequential} pattern, \athena\ prefetches the next $N$ (defaulted to 4) sequential items following the current semantic or indexical order.
For the \emph{random} pattern, \athena\ adopts a \emph{statistical prefetching} strategy, where the entire dataset is prefetched into the cache if the expected cache hit ratio is over a threshold.
For the \emph{skewed} pattern, prefetching is less effective and thus \athena\ opts not to prefetch at all.
\end{sloppypar}

\begin{figure}
    \centering
    \includegraphics[width=0.4\textwidth]{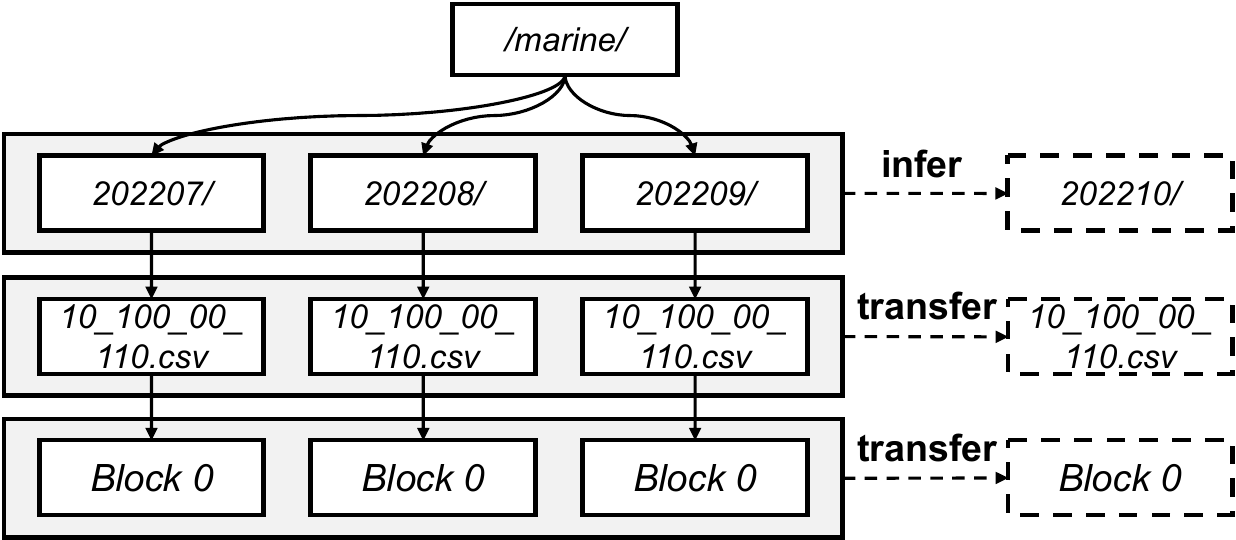}
    \caption{Example of hierarchical prefetching over an ocean atmosphere dataset (ICOADS~\cite{cisl_rda_dsd548000}), with semantic information in path names. The directory \texttt{202209/} contains all data collected in September 2022 (a total of 10 csv files), and the file \texttt{10\_100\_00\_110.csv} stores the data for a specific geographical region (longitude between $0$ and $10$, and latitude between $100$ and $110$). 
        Consider an AI workload that performs inference using all historical data for the location described in \texttt{10\_100\_00\_110.csv}. In this case, a sequential pattern can be identified at the directory level, yielding a prefetching candidate of directory \texttt{202210/}.
        Since only \texttt{10\_100\_00\_110.csv} is visited in prior sibling nodes, under \texttt{202210/} we only prefetch the \texttt{10\_100\_00\_110.csv} file. 
        Such selective prefetching can also be applied at the block level.
    }
    \label{fig:hierarchical_prefetch}
\end{figure}

\begin{sloppypar}
Regarding granularity adaptivity, 
we devise a technique called \emph{hierarchical prefetching} to support data prefetching at any arbitrary granularity.
In the horizontal direction, we first apply the previous principles to determine the prefetching candidates at the current granularity. 
For non-leaf \texttt{AccessStream} nodes which correspond to abstracted groups of low-level accesses,
we need to vertically complement the prefetching strategy. 
In particular, 
some blocks or files (e.g., metadata files) may be accessed more frequently than others in the same path. To account for this, we selectively prefetch the hot units for better cache efficiency.
For instance, if $a_n$ represents the $n$-th file access and block 0 appears $x$ times among the records from $a_1$ to $a_n$, the access probability $f$ can be calculated as $f = x/n$.
Given the threshold $f_p$, items with a probability $f$ above $f_p$ would be considered for prefetching.
Such a relative position would then be enforced when prefetching the subsequent files.
\Cref{fig:hierarchical_prefetch} shows a realistic example of hierarchical prefetching.
\end{sloppypar}

\begin{sloppypar}
\phm{Eviction Strategy Optimization.} 
\athena\ adaptively switches the eviction policy based on the access pattern of each \texttt{AccessStream}.
To be specific, for sequential patterns, it adopts eager eviction, i.e., immediately evicting a sample after it is accessed.
Since each accessed item is typically accessed only once, there is no need to cache these items\footnote{
    That said, the caching frameworks can still accept explicit user instructions to persistently cache (or never cache) a dataset.
    Our primary objective here is to provide an adaptive strategy that can behave well in general cases.
}. 
For random patterns, the uniform eviction strategy (\Cref{sec:prior}) is employed to prevent cache thrashing. 
For skewed access patterns, the LRU eviction strategy is used as the general solution. 
Besides, the eviction strategies of \athena\ are also general to the specific storage granularity. The eviction unit can be at the block, file, or even directory level, depending on where a non-trivial data access pattern exists.
\end{sloppypar}

Moreover, caching frameworks like Alluxio employ a \emph{Time-To-Live} (TTL) hyperparameter to enable proactive eviction (\Cref{sec:prior}). However, setting TTL values correctly is challenging in practice~\cite{yang2021large}.
In \athena\ we propose an adaptive method to automatically set the TTL for \texttt{AccessStreams} with random access patterns. The key observation is that the temporal gap between consecutive accesses tends to follow a \emph{normal} distribution. 
For each non-trivial \texttt{AccessStream}, we can collect the temporal gaps between accesses in the observation window and then fit the distribution to estimate the mean value and the standard deviation. 
Based on this, we can set the TTL value to the maximum time gap under a given significance threshold.
If the time elapsed since the last access of an \texttt{AccessStream} is larger than the TTL, it is unlikely that a new request under that \texttt{AccessStream} would come later. This suggests that the corresponding job has likely finished. To prevent unused data from occupying cache space, \athena\ evicts the entire dataset associated with that job, making the cache space available for other active workloads.
To prevent accidental eviction due to small temporal disturbances, a small base time is added to the TTL to ensure that only genuinely idle data is evicted. This approach provides timely eviction of completed jobs, maximizing cache utilization for active tasks.

\phm{Cache Allocation Optimization.}
Cache space is usually an expensive resource competed by different tenants.
As explained in \Cref{sec:prior}, both static allocation and shared allocation without isolation are inefficient. We need to properly allocate the cache space to different tenants (i.e., to the non-trivial \texttt{AccessStreams}) for better efficiency.
To achieve this, we introduce a metric, denoted by $\mathcal{B}$, to quantify the marginal cache demand of each workload.
This metric represents \emph{the transmission amount that can be reduced in unit time by allocating an additional unit of cache space to a workload}.

The estimation of $\mathcal{B}$ depends on the access pattern of the workload. For a data item just accessed (with a volume size of $s$), if we have the information of how long ($t$) it will take before the same data is accessed again, the marginal benefit can be expressed by $\mathcal{B} = \frac{s/t}{s} = 1/t$.
For \emph{sequential} patterns, once a data item is accessed, it will not be accessed again. Hence, the marginal benefit is $\mathcal{B} = 0$.
For \emph{random} patterns, each data item is accessed exactly once in a given epoch. Suppose the dataset contains $n$ block items and the inter-access temporal gap is $g$, we have $t = g \cdot n$ and $\mathcal{B} = 1/(g \cdot n)$.
Note that this method can naturally apply to the case where multiple training jobs are sharing the same datasets ($g$ proportionally smaller).
For \emph{skewed} patterns, we instead adopt a sampling method inspired by the classical ``ghost cache'' method~\cite{bansal2004car,megiddo2003arc}, which captures the evicted data items that could have been a hit if the cache size enlarges. 
To be specific, for each item that is evicted from the cache, we add it to a data structure called \emph{BufferWindow}
(containing at most $w$ block items), which uses the same eviction algorithm as the cache.
For each item requested, if it is found in the BufferWindow, its remote-fetching overhead can potentially be eliminated if we increase the cache allocation by the BufferWindow size. 
Thus, we can measure the frequency of such BufferWindow hit (denoted by $f_\text{BufferHit}$), and the marginal benefit can be directly quantified by $\mathcal{B} = \lambda \cdot f_\text{BufferHit}/w$, where $\lambda$ is the request arrival rate.

With such a uniform metric, \athena\ can dynamically adjust cache resources between workloads for better efficiency. 
Moreover, cache space is isolated at the \texttt{AccessStream} level. If the occupied space of an \texttt{AccessStream} exceeds its quota, cache eviction is enforced on the items under that \texttt{AccessStream}, ensuring that other workloads are not negatively impacted by cache overuse.			

\section{Implementation}
\label{sec:implementation}

We have implemented \athena\ atop JuiceFS~\cite{JuiceFS}, an open-source distributed file system that serves as a shared cache for cloud storage.
We integrate \athena\ as a pluggable module, with approximately 5,000 lines of Go code.
\athena\ is composed of two components: \texttt{AccessStreamTree} for status monitoring and \texttt{CacheManageUnit} for action enforcement.


\phm{Status Monitoring.}
To monitor the data request status, we add a hook in FUSE \texttt{Read()} interface, which reports I/O request information (e.g., file inode, offset, and I/O size) to the \texttt{AccessStreamTree}. 
Accesses are added to a proper level in the \texttt{AccessStreamTree} as described in \Cref{sec:AccessStreamTree}. 
For an \texttt{AccessStream}, once its number of child nodes exceeds the observation window size (set to 100), the node is considered non-trivial, triggering the pattern recognition in \Cref{sec:diagnosis}. For the K-S test, we use the \texttt{kstest()} function in the \texttt{scipy.stats} library.
Based on the detected access pattern, the \texttt{AccessStream} generates prefetching and eviction candidates, and calculates the marginal benefit, all based on the customized policies elaborated in \Cref{sec:adaptive_management}. 
With such information, we create a \texttt{CacheManageUnit} object that manages the cache space for the corresponding \texttt{AccessStream}.




\phm{Action Enforcement.}
To adaptively manage the shared cache in a fine-grained manner, we maintain a series of \texttt{CacheManageUnit} each mapped to an \texttt{AccessStream} to enforce the customized caching policies and ensure \texttt{AccessStream}-level cache isolation. 
\texttt{CacheManageUnit} calls the \texttt{vfs.read()} API to read prefetching candidates and the \texttt{os.remove()} API to evict candidates. 
The cache space is shifted between different \texttt{CacheManageUnits} based on the marginal benefit. 
The shift is conducted in rounds (every 60s) and the per-round shifting amount is 640 MB. Each dataset remains a minimum cache share despite with such cache migration. Whenever the cache size of an \texttt{AccessStream} changes, \athena\ refreshes its access pattern, policies, and marginal caching benefit for accurate and timely status monitoring.

\phm{Overhead Control.}
To mitigate the management overhead of \athena, we adopt a series of methods.
First, we adopt \emph{layer compression} and \emph{node pruning} techniques.
Layer compression merges trivial \texttt{AccessStream} nodes into a single \texttt{AccessStream} node in deep paths with many non-bifurcating directory layers.  
Node pruning removes earlier child nodes when a non-trivial \texttt{AccessStream} node exceeds the observation window size threshold. 
We also set a hard limit (10,000) to the maximum number of nodes in the \texttt{AccessStreamTree}, removing excess nodes using LRU. 
Meanwhile, the pattern and policy analysis in each \texttt{AccessStream} run in a separate Go routine in parallel with the regular request serving logic, without stalling the critical path. 
Regarding the complexity, let \( N \) denote the number of nodes in \texttt{AccessStreamTree}; 
since the child number of each node is bounded by the observation window size,  the time complexity of \athena \ for searching and updating the tree during an I/O request is is \( O(\log N) \), and the space complexity of \athena \  is \( O(N) \).
\section{Evaluation}
\label{sec:evaluation}


\subsection{End-To-End Performance}
\label{sec:eval_macro}

\phm{Hardware Setup.} 
In our experiments, we adopt a hardware setup with storage-compute disaggregation. 
The local computing cluster includes 6 A100 GPUs and 12 V100 GPUs in our laboratory environment. 
Meanwhile, we maintain a dedicated local cache server\footnote{
The community version of JuiceFS currently does not support deploying multiple cache nodes, yet we note that this does not affect the validity of our \athena \ experiments: distributed cache servers, if deployed, still work as a single cache pool, and the intra-cluster communication cost is negligible when compared to the remote access cost to S3.
} with 4 Intel 6133 CPUs, 256 GB DRAM, and 1 TB SSD.
It provides caching services to the computing infrastructure through NFS, connects to AWS S3~\cite{S3} for remote data storage, and additionally maintains a shared distributed cache by connecting client hosts’ disk cache via NFS.
Our measurements show that the average remote data fetching bandwidth is around 1 Gbps and the delay is around 150 ms.

\phm{Workloads.} 
We create a diverse workload suite as summarized in \Cref{tab:workloads_patterns}. 
There are totally 18 different workloads involving various patterns. 
For clarity, we assign each workload with a static job id.
The available shared cache size is set to $150$ GB, which accounts for approximately 35\% of the total dataset size.
The job submission gaps follow a Poisson distribution~\cite{ren2013workload,zhang2024cross} with an expected interval of $\beta=60$ s. 

\begin{table}
\centering
\scriptsize
\begin{tabular}{|>{\centering\arraybackslash}p{1.9cm}|>{\raggedright\arraybackslash}p{3.8cm}|>{\centering\arraybackslash}p{1.8cm}|}
\hline
\textbf{Dataset} & \textbf{Workload \& Model} & \textbf{Access Pattern} \\ \hline
\multirow{2}{*}{AudioMNIST~\cite{becker2024audiomnist}} 
    & \scriptsize{\textcircled{\tiny{1}}}\scriptsize\  [V] VGG16~\cite{simonyan2014very} Training (Dataset Shuffled in Memory) & \multirow{2}{*}{Sequential} \\ \hline
\multirow{1}{*}{FashionProduct~\cite{param_aggarwal_2019}} 
    & \scriptsize{\textcircled{\tiny{2}}}\scriptsize\ [V] VGG16 Test & \multirow{1}{*}{Sequential} \\ \hline
\multirow{1}{*}{AirQuality~\cite{noauthor_time_nodate}} 
    &\scriptsize{\textcircled{\tiny{3}}}\scriptsize\ [C] Air Quality Analysis & \multirow{1}{*}{Sequential} \\ \hline
\multirow{1}{*}{ICOADS~\cite{cisl_rda_dsd548000}} 
    & \scriptsize{\textcircled{\tiny{4}}}\scriptsize\ [C] Marine Data Analysis & \multirow{1}{*}{Sequential} \\ \hline
\multirow{3}{*}{Bookcorpus~\cite{zhu2015aligning}} 
    & \scriptsize{\textcircled{\tiny{5}}}\scriptsize\ [C] Data Preprocessing & Sequential \\ \cline{2-3}
    &\scriptsize{\textcircled{\tiny{6}}}\scriptsize\ [A] OPT-125M~\cite{zhang2022opt} Checkpoint Loading & \multirow{2}{*}{Sequential} \\ \cline{2-3}
    & \scriptsize{\textcircled{\tiny{7}}}\scriptsize\ [A] OPT-125M~\cite{zhang2022opt} Finetuning & Random \\ \hline
\multirow{3}{*}{ImageNet~\cite{deng2009imagenet}}
    & \scriptsize{\textcircled{\tiny{8}}}\scriptsize\ [V] ResNet-50~\cite{he2016deep} Test & Sequential \\ \cline{2-3}
    & \scriptsize{\textcircled{\tiny{9}}}\scriptsize\ [V] ResNet-50 Training & Random \\ \cline{2-3}
    & \scriptsize{\textcircled{\tiny{10}}}\scriptsize\ [V] AlexNet~\cite{krizhevsky2012imagenet} Training & Random \\ \hline
\multirow{3}{*}{MITPlaces~\cite{zhou2014learning}} 
    & \scriptsize{\textcircled{\tiny{11}}}\scriptsize\ [V] AlexNet Test & Sequential \\ \cline{2-3}
    & \scriptsize{\textcircled{\tiny{12}}}\scriptsize\ [V] ResNet-50 Training & Random \\ \cline{2-3}
    & \scriptsize{\textcircled{\tiny{13}}}\scriptsize\ [V] AlexNet Training & Random \\ \hline
\multirow{2}{*}{LakeBench~\cite{deng2024lakebench}} 
    & \scriptsize{\textcircled{\tiny{14}}}\scriptsize\ [C] Table Join & Skewed \\ \cline{2-3}
    & \scriptsize{\textcircled{\tiny{15}}}\scriptsize\ [C] Table Union & Skewed \\ \hline
\multirow{2}{*}{Wiki~\cite{wiki_rag}} 
     & \scriptsize{\textcircled{\tiny{16}}}\scriptsize\ [V] RAG (Large) & Skewed \\ \cline{2-3}
    & \scriptsize{\textcircled{\tiny{17}}}\scriptsize\ [V] RAG (Small) & Skewed \\ \hline
\multirow{2}{*}{LLaVa Dataset~\cite{liu2023llava}} 
    & \scriptsize{\textcircled{\tiny{18}}}\scriptsize\ [4*A] Multi-modal Finetuning (Text~\cite{liu2023llava} and Images~\cite{lin2014microsoft}) & Sequential and Random \\ \hline
\end{tabular}
\caption{Workloads in our experiments. The character [A], [V], [C] respectively represents that the job is executed on A100 GPU, V100 GPU and purely on CPU.}
\label{tab:workloads_patterns}
\vspace{-0.15in}
\end{table}

\phm{Baselines.} 
For end-to-end evaluation, we compare \athena\ performance with the vanilla JuiceFS as well as the case without caching.
JuiceFS adopts block-level prefetching for sequential access patterns and an LRU-like eviction strategy.
Note that the focus of this paper is not to propose new caching methods for certain workload types, but to unleash the power of existing workload-specific caching methods by enabling adaptivity. 
Therefore, the demonstrated adaptivity benefit here also applies to other emerging workload-specific strategies.
Besides, another popular caching framework, Alluxio~\cite{Alluxio}, shares almost the same cache management policies as JuiceFS, thus our experimental conclusions also apply to Alluxio customizations. 
We will respectively compare \athena\ with more baselines in later micro-benchmark experiments. 

Regarding the hyperparameter setup of \athena, for pattern recognition (\Cref{sec:diagnosis}), we set the significance level $\alpha$ to 0.01.
Meanwhile, for policy customization (\Cref{sec:adaptive_management}), we set the prefetching threshold $f_p$ to 0.8 and the prefetch depth to 4. We configure the BufferWindow size $w$ to 100.

\phm{Metrics.}
To evaluate the caching effectiveness, we measure two metrics in each case.
The first is the average Job Completion Time (JCT), and the second is the overall Cache Hit Ratio (CHR), which is the percentage of (block-level) data accesses that are served directly by the local cache rather than by the remote storage.

\begin{figure}
    \centering
    \includegraphics[width=0.45\textwidth]{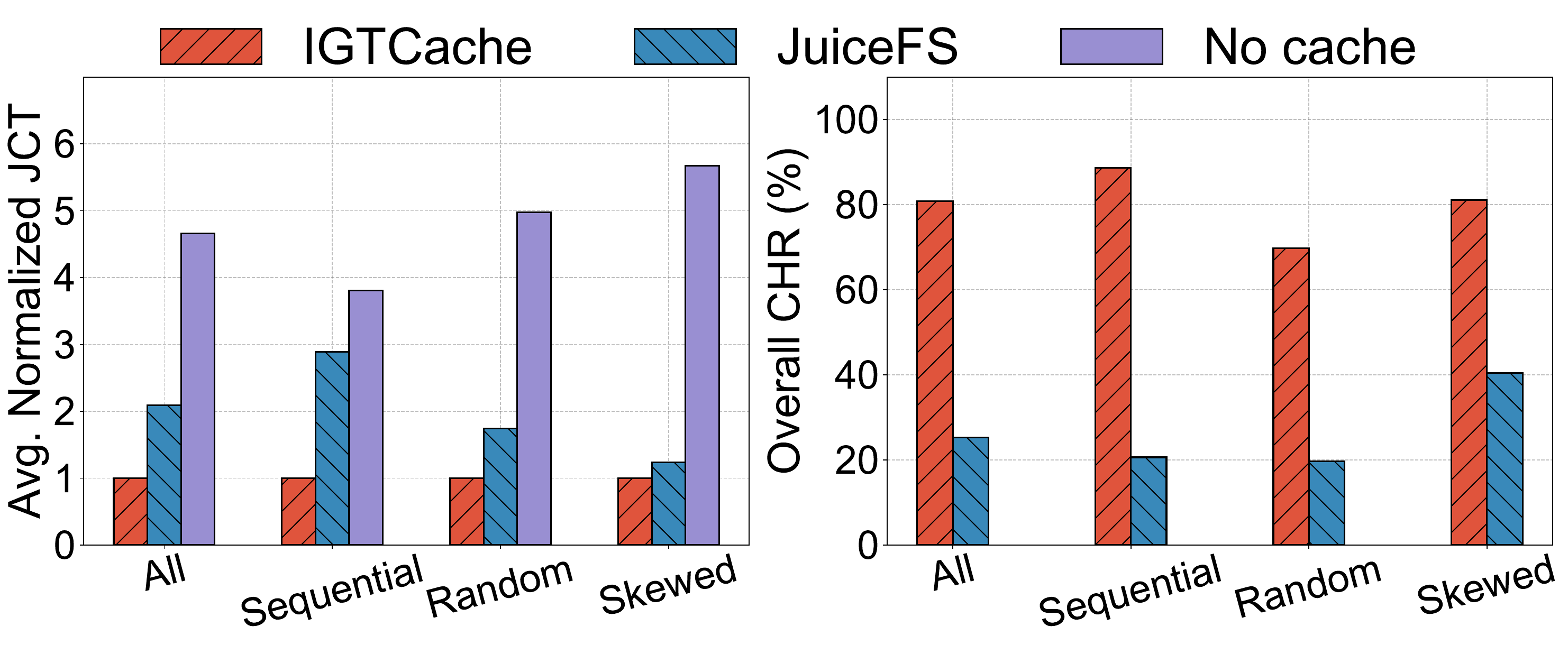}
    \vspace{-0.1in}
    \caption{JCT and CHR in end-to-end evaluations.}
    \label{overall}
    \vspace{-0.1in}
\end{figure}

\phm{Overall Performance.}
\Cref{overall} shows the average JCT (normalized by that of \athena) and CHR of all the jobs in \Cref{tab:workloads_patterns}, including the performance for each pattern subset.
According to \Cref{overall}, we observe that caching frameworks are indispensable for scenarios with compute-storage disaggregation. Specifically, the average JCT with JuiceFS cache is 55.0\% better than that without cache.
Furthermore, \athena\ significantly surpasses existing caching practices. Compared with JuiceFS, \athena\ reduces the average JCT by 52.2\% and increases the overall CHR by 55.6\%.
In particular, \athena\ effectively improves job performance for all workload patterns, thus confirming its generality.

Next we resort to micro-benchmark experiments to check the benefit of \athena\ in each cache management aspect.

\subsection{Prefetching Performance}
\label{sec:eval_prefetch}

\begin{figure}
\centering
    {\includegraphics[width=0.43\textwidth]{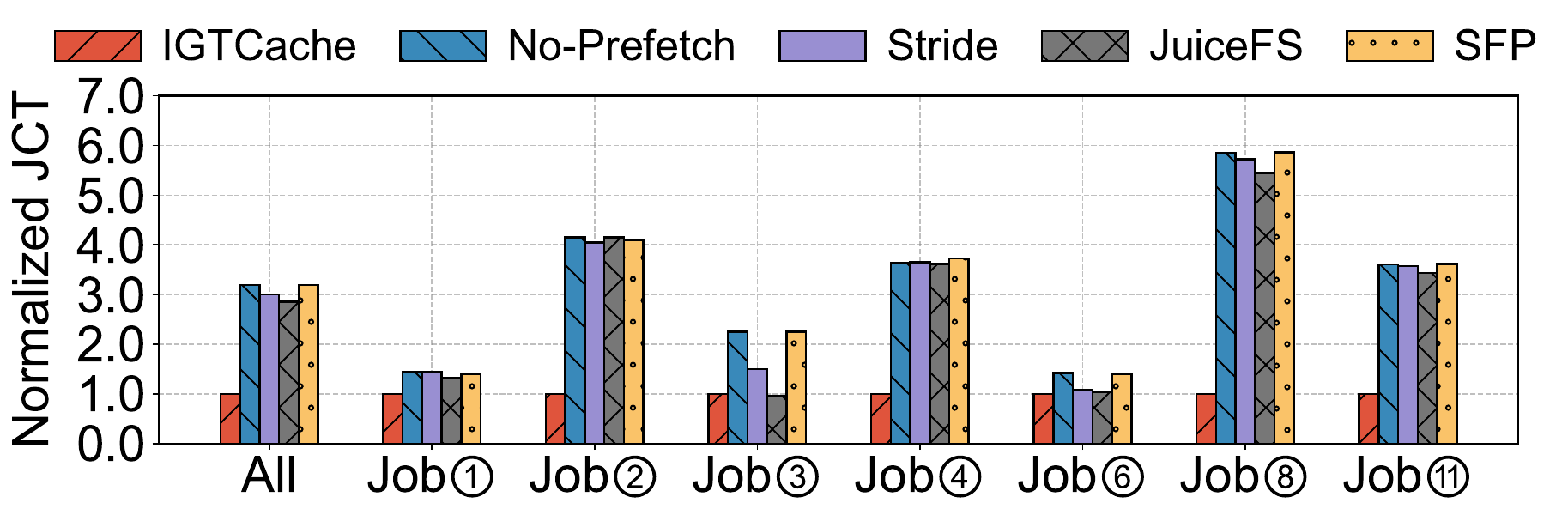}}
    \hspace{0.00\textwidth} 
    {\includegraphics[width=0.43\textwidth]{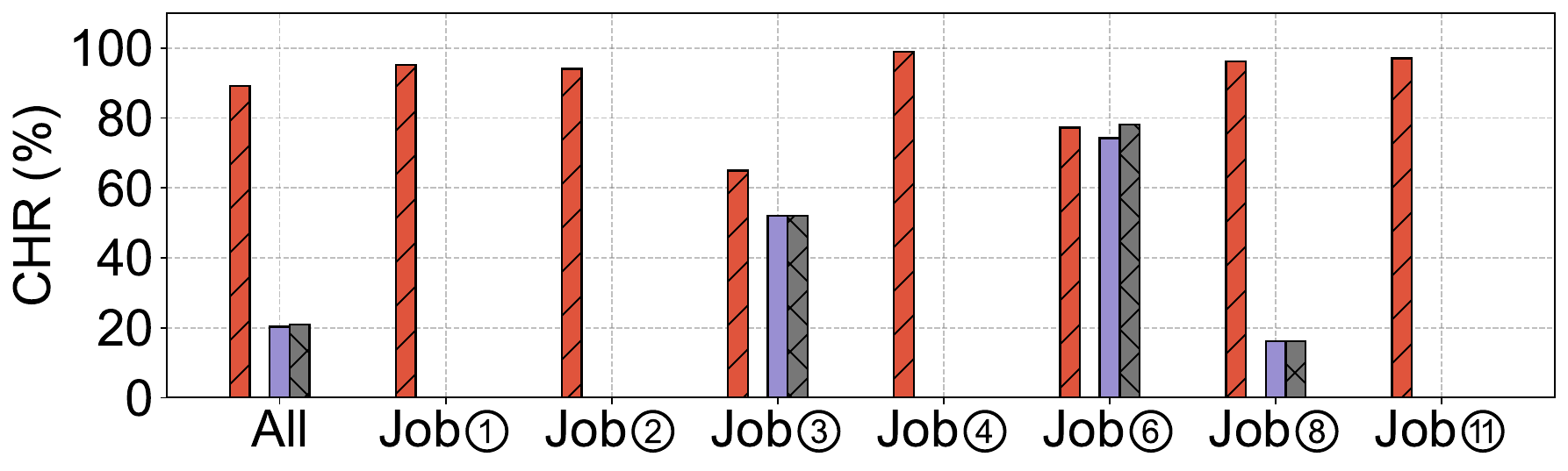}}
    \vspace{-0.1in}
\caption{Performance of different prefetching schemes.}
\label{prefetching}
\vspace{-0.1in}
\end{figure}

\phm{Setup.}
To evaluate \athena\ performance in prefetching, we select a subset of jobs that are sensitive to the prefetching effect (job IDs marked in \Cref{prefetching}).
Meanwhile, we disable all the \athena's functionalities except prefetching and compare against four baselines: 
1) stride prefetching: it prefetches 4 additional sequential blocks after every 4 consecutive block accesses,
2) enhanced stride prefetching~\cite{juicefs-cache}: the default scheme in JuiceFS which adaptively increases the block prefetching depth if with strong consecutiveness,
3) SFP~\cite{wang2021sfp}: a file-level prefetching strategy leveraging the historical associations between files expressed in a Markov chain,
and 4) no prefetching.
We record the overall JCT and CHR for each selected job. The results are shown in \Cref{prefetching}. 

\phm{Results.} 
According to \Cref{prefetching}, \athena\ achieves the best caching performance in all cases, attaining a JCT reduction of 64.9\% and a CHR improvement of 68.2\% compared to the second-best method. 
Specifically, for sequential block reading workloads (like job-\normalsize{\textcircled{\scriptsize{6}}}\normalsize), the performance of \athena\ is comparable to the stride-prefetching methods. However, for other workloads with sequential file reading, the performance of \athena\ significantly outperforms the alternatives.
This is because these datasets are composed of massive small files, where the file size is comparable to or even smaller than a block (4MB). 
In this case, block-level prefetching is largely ineffective.
Meanwhile, different from traditional scenarios~\cite{kroeger2001design,wang2021sfp}, AI datasets typically have few files that are strongly correlated or accessed repeatedly, rendering association-based file prefetching~\cite{kroeger2001design} ineffective.

Moreover, we verify the effectiveness of the hierarchical prefetching technique. For job-\normalsize{\textcircled{\scriptsize{4}}}\normalsize, which, as elaborated in \Cref{prefetching}, traverses the atmosphere data of a given location (each represented as a file) across different months (each represented as a directory), hierarchical prefetching reduces the JCT by 64.4\%. Without hierarchical prefetching, where all the files in the directory-to-prefetch are fetched, the I/O would soon be exhausted, causing a JCT inflation of 15.7$\times$. We also evaluate the statistical prefetching technique designed for random pattern. We compare the completion time of job-\normalsize{\textcircled{\scriptsize{7}}}\normalsize \ with and without dataset prefetching enabled. Our observations show that enabling dataset prefetching reduces the completion time of the first epoch by 6.8\%. 

\subsection{Eviction Performance}
\label{sec:eval_eviction}

\begin{figure}
\centering
    {\includegraphics[width=0.43\textwidth]{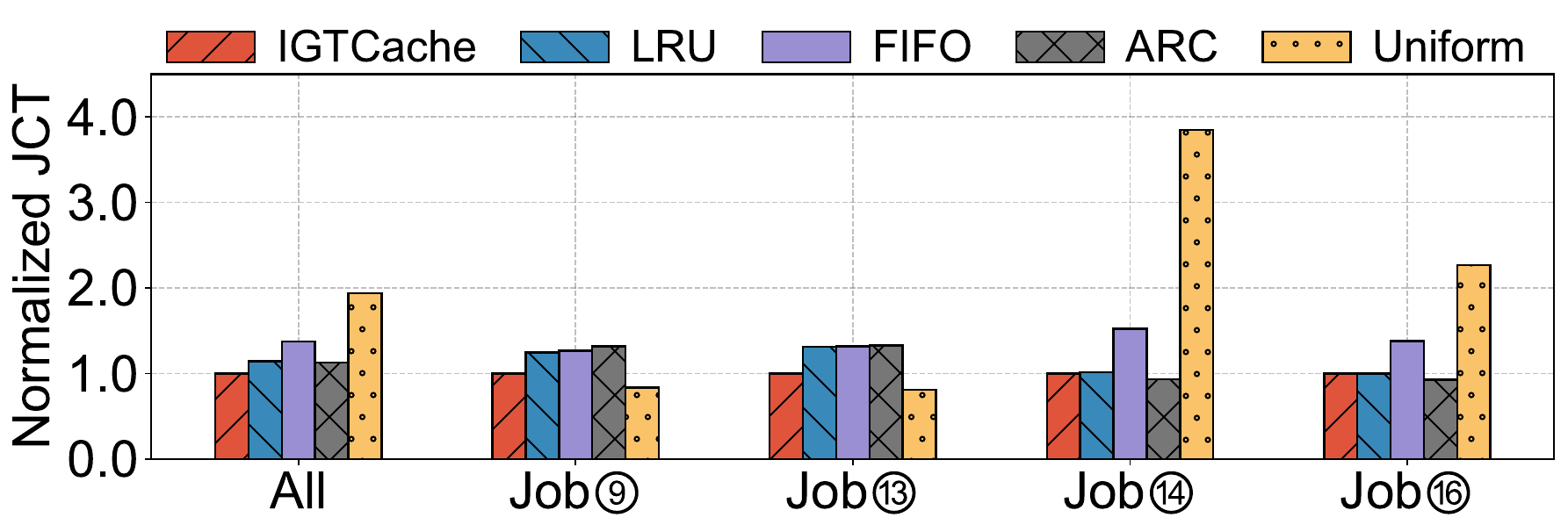}}
    \hspace{0.00\textwidth} 
    {\includegraphics[width=0.43\textwidth]{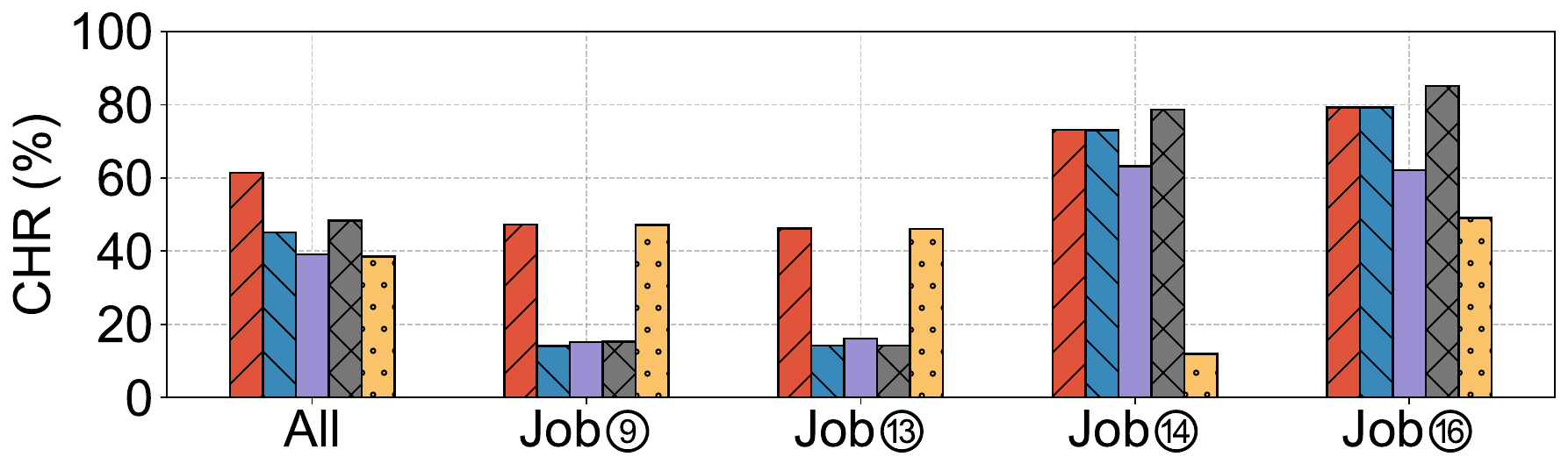}}
\vspace{-0.15in}
\caption{Performance of different eviction schemes.}
\label{eviction}
\vspace{-0.15in}
\end{figure}

\phm{Setup.} 
To evaluate the eviction performance of \athena, we select a set of eviction-sensitive workloads, involving both random and skewed patterns. 
{We set the cache size for each job to 50\% of its dataset size, as prior researches~\cite{mohan2020analyzing} have shown that this configuration well reflects the performance of cache eviction strategies with AI workloads.}
We disable all \athena\ functionalities other than eviction, and compare it against four baselines: 1) LRU, 2) FIFO, 3) ARC~\cite{megiddo2003arc}---an adaptive eviction strategy that balances recency and frequency for traditional workloads, and 4) uniform caching~\cite{zhao2023silod}.
\Cref{eviction} shows the normalized JCT and CHR. 

\phm{Results.}
According to \Cref{eviction}, \athena\ consistently achieves the best performance in all cases. 
Specifically, it adopts uniform caching for training workloads and LRU for query workloads, thus doing well in both sides. 
Compared to the second-best strategy, \athena\ reduces the average JCT by 11.2\% and increases the overall CHR by 13.2\%. 
We observe that although ARC claims to be an adaptive replacement strategy, its performance degrades to the same level as LRU on random patterns (job-\normalsize{\textcircled{\scriptsize{9}}}\normalsize \ and job-\normalsize{\textcircled{\scriptsize{13}}}\normalsize).

\begin{figure}
    \centering
    \includegraphics[width=0.38\textwidth]{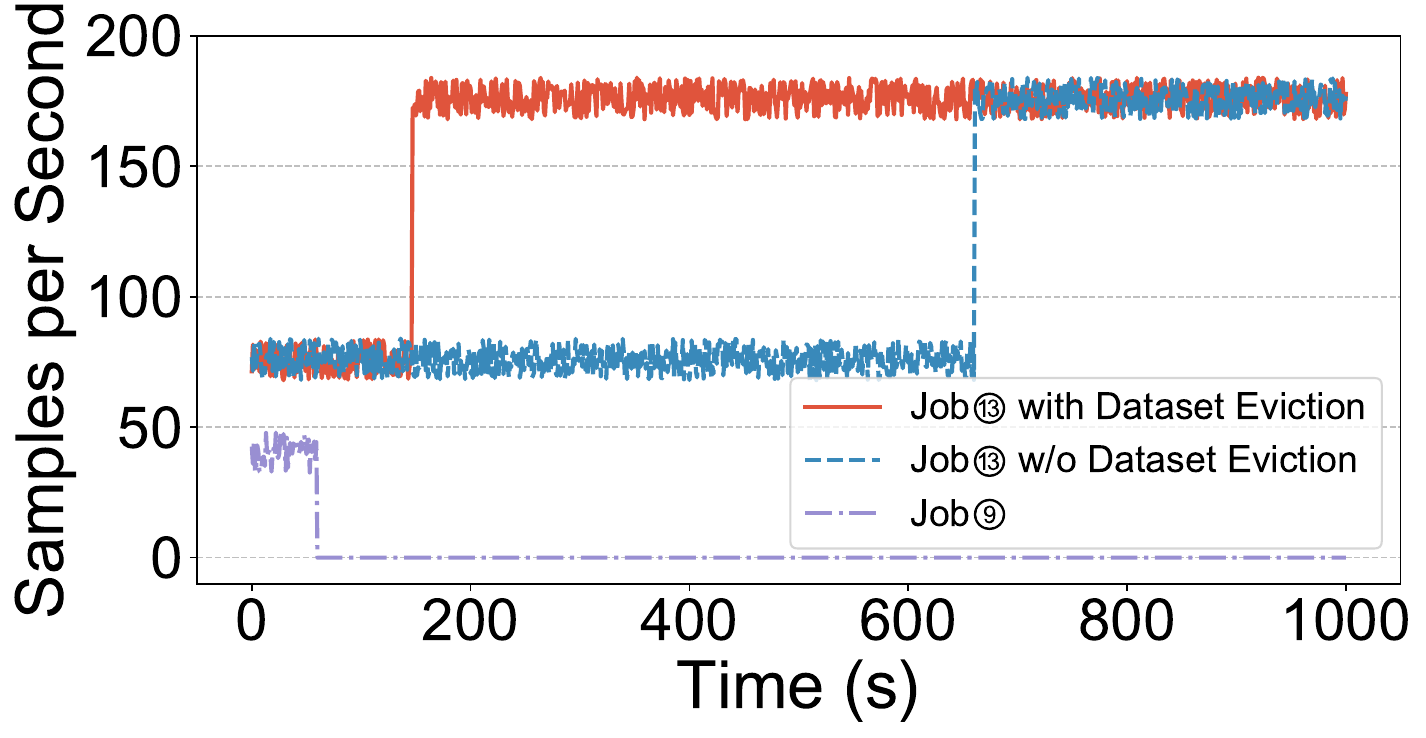}
    \vspace{-0.1in}
    \caption{After job-\normalsize{\textcircled{\scriptsize{9}}}\normalsize \ finishes at 60s, the eviction of its ImageNet dataset starts at 146s if using adaptive TTL setup in \athena, and at 660s if using default TTL setup.}
    \label{fig:dataset_eviction}
    \vspace{-0.15in}
\end{figure}
We also verify the effectiveness of adaptive TTL setup (\Cref{sec:adaptive_management}) under \athena.
We simultaneously launch two model training jobs: job-\normalsize{\textcircled{\scriptsize{9}}}\normalsize \
and job-\normalsize{\textcircled{\scriptsize{13}}}\normalsize,
which evenly share the cache space.
At the time of 60s, we manually stop job-\normalsize{\textcircled{\scriptsize{9}}}\normalsize \ (with ImageNet dataset), and measure the throughput of job-\normalsize{\textcircled{\scriptsize{13}}}\normalsize \ respectively with the default\footnote{
Such a TTL value appears in hyper-parameter setup instructions of those frameworks and may change in different cases~\cite{yang2021large}. 
Our experiment here is to show that it is inconvenient and also risky to arbitrarily set the TTL value, and workload-adaptive TTL setup is a promising direction to explore.
}
TTL setup in JuiceFS and the adaptive TTL setup in \athena.
\Cref{fig:dataset_eviction} shows the instantaneous throughput of both jobs. 
With JuiceFS, the eviction of the ImageNet dataset occurs only after 600s. In contrast, 
under \athena, 
with the significance threshold $0.01$, 
\athena\ automatically sets the TTL to 86s (with a base time of 60s).
Such a shorter TTL allows \athena\ to release the cache space occupied by ImageNet earlier, making it available for MITPlaces, thus achieving a higher cache utilization as well as a higher throughput.

\subsection{Cache Capacity Allocation Performance}
\label{sec:eval_allocation}
\vspace{-0.1in}
\phm{Setup.}
To evaluate \athena\ performance in cache allocation optimization, we select four jobs sensitive to cache space (two training jobs---job-\normalsize{\textcircled{\scriptsize{9}} and job-\normalsize{\textcircled{\scriptsize{13}}, and two query jobs---job-\normalsize{\textcircled{\scriptsize{14}} and job-\normalsize{\textcircled{\scriptsize{16}}).
To simplify the experiment, we scale down the dataset size of the four selected jobs by $10 \times$. Accordingly, we also scale down the shared cache size to $7.5$ GB.
We compare \athena\ with three baselines: JuiceFS~\cite{JuiceFS}, Quiver~\cite{kumar2020quiver}, and Fluid~\cite{gu2022fluid}.
JuiceFS lets each job freely use the shared cache without isolation.
Quiver directly profiles the caching benefit of each training job and allocates cache resources to the job with higher benefits.
Fluid determines the cache size of each training job proportionally to the total batch size. 
Since both Quiver and Fluid are designed only for model training workloads, we extend them to handle mixed workloads. 
For Quiver, we evenly divide the cache space between the two workload types.
For Fluid, all the cache space not claimed by model training jobs is allocated to query workloads.  

\phm{Results.}
As shown in \Cref{fig:eval_allocation}, \athena\ attains the best performance in both overall JCT and CHR. Compared to the second-best strategy, \athena\ reduces the average JCT by 7.5\% and increases the overall CHR by 10.1\%. 
Regarding the reasons behind this, we depict the instantaneous marginal cache benefit and cache allocation amount for each job in \Cref{fig:eval_marginal}
(the minimum share of each job is set to 640MB).
As shown in \Cref{fig:eval_marginal}, 
the marginal benefits of different workloads are indeed dynamic, and it is thus necessary to enable runtime cache shifting.
In contrast, Quiver and Fluid do not support shifting cache resources between workloads with different access patterns; JuiceFS does not support workload-level cache isolation, and query workloads with temporarily intense requests may edge out the training dataset.

\begin{figure}[t]
\centering
    {\includegraphics[width=0.43\textwidth]{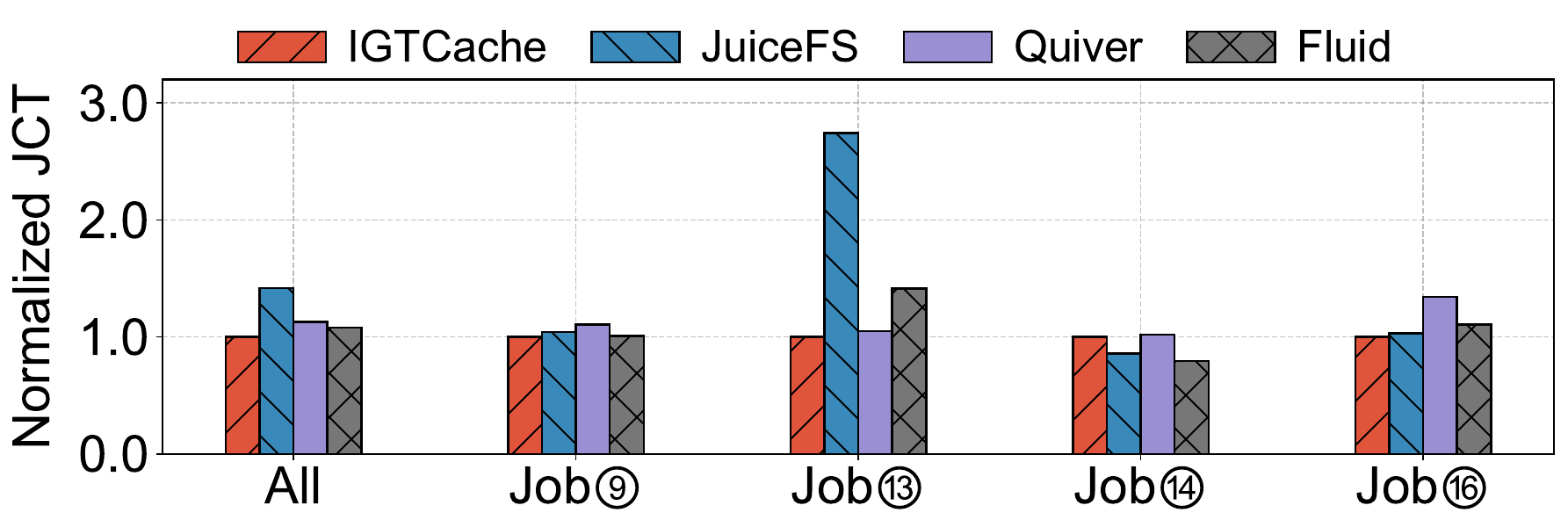}}
    \hspace{0.00\textwidth} 
    {\includegraphics[width=0.43\textwidth]{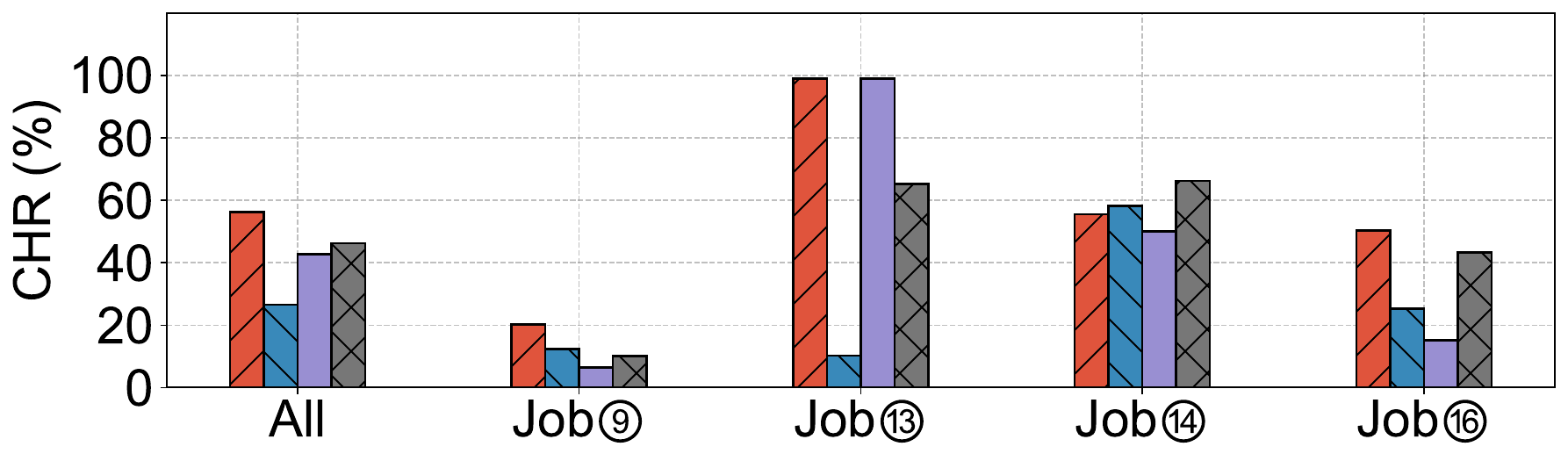}}
    \vspace{-0.1in}
    \caption{Performance of different allocation schemes.}
\label{fig:eval_allocation}
\vspace{-0.2in}
\end{figure}


\begin{figure}
\centering
    {\includegraphics[width=0.43\textwidth]{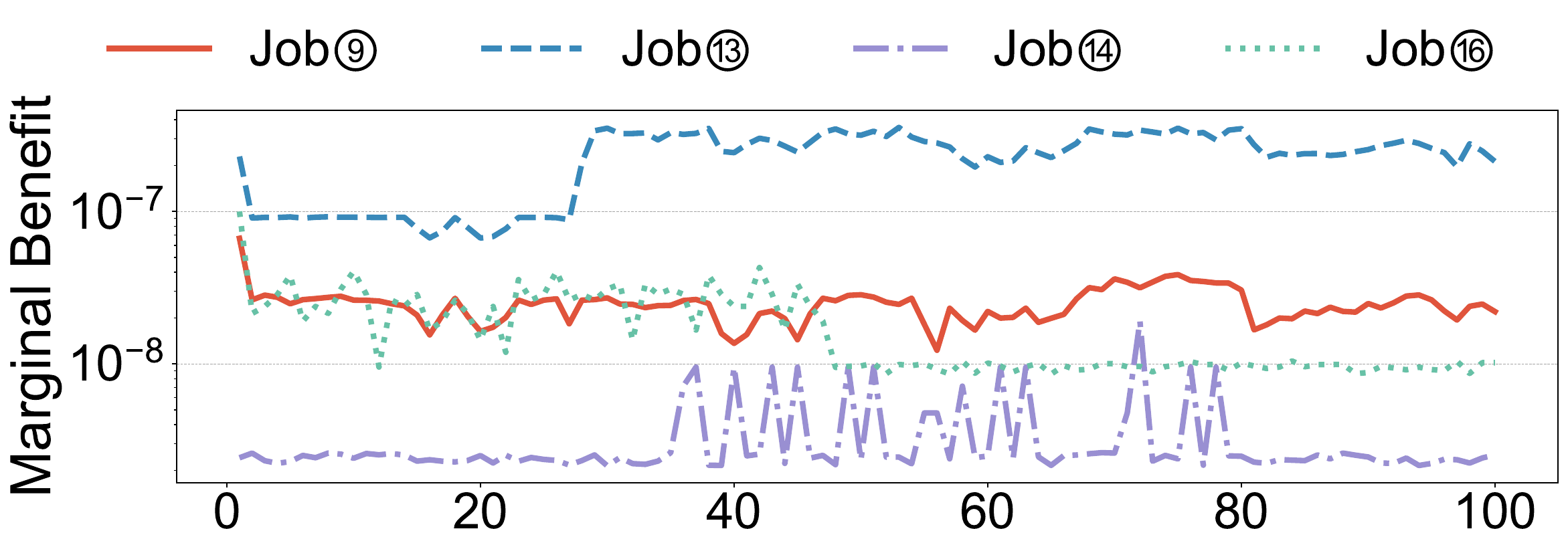}}
    \hspace{0.00\textwidth} 
    {\includegraphics[width=0.43\textwidth]{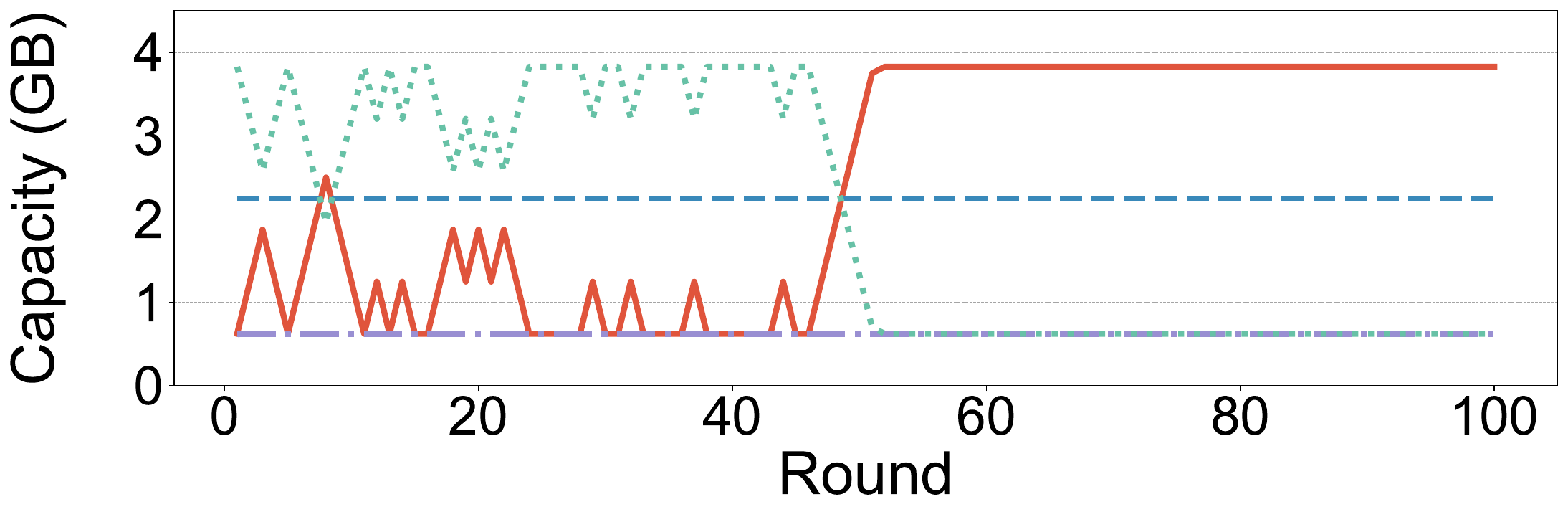}}
    \vspace{-0.1in}
\caption{Instantaneous marginal cache benefit and cache allocation amount of each job under \athena{}.}
\label{fig:eval_marginal}
\vspace{-0.05in}
\end{figure}

\subsection{Sensitivity and Overhead Analysis}
\label{sec:eval_overheads}


\phm{Impact of K-S test parameters.}We examine the impact of the significance level $\alpha$ used in the K-S test on the pattern prediction accuracy, as discussed in \Cref{sec:diagnosis}. 
For the purpose of this analysis, four workloads are selected from \Cref{tab:workloads_patterns} for both the random pattern and the skewed pattern.
Using an observation window of size 100, we collect access sequences and perform the K-S test to identify the patterns. Each workload is evaluated 100 times (with different access sequences) to assess the pattern recognition accuracy under each p-value. As shown in \Cref{accuracy_alpha}, no $\alpha$ can always yield the best performance yet the value of 0.01 is in general better than 0.05; in fact, both are sufficiently good. 


Moreover, we fix the significance level $\alpha$ at 0.01 and evaluate the impact of different observation window sizes on pattern recognition accuracy. As shown in Figure 16, a small window size (e.g., 10) leads to poor prediction accuracy due to insufficient samples. Increasing the size to 100 significantly improves accuracy, while further increasing to 1000 yields little additional benefit. Thus, we set the default observation window size to 100, balancing accuracy and overhead.



\begin{figure}
  \centering
  \begin{minipage}[b]{0.23\textwidth}
    \centering
    \includegraphics[width=\textwidth]{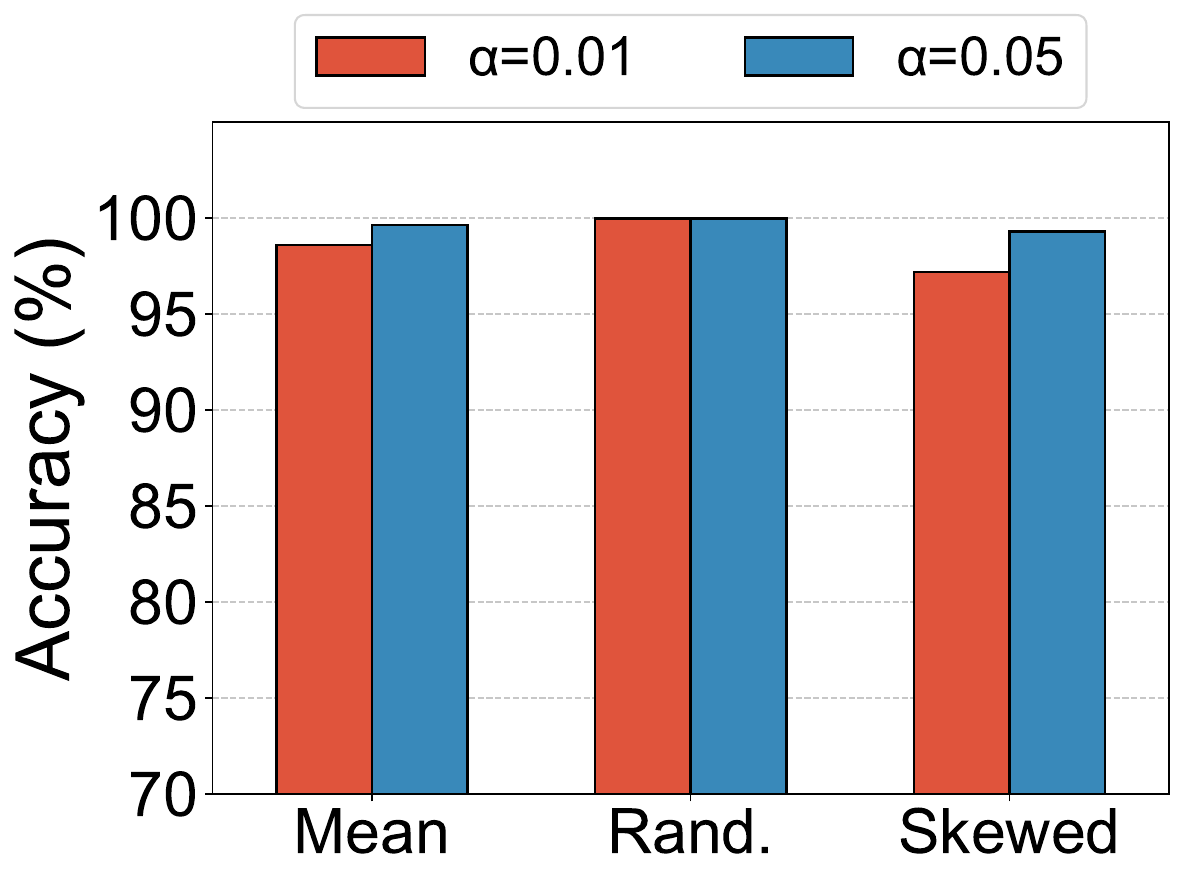}
    \captionof{figure}{Accuracy with different confidence $\alpha$.}
    \label{accuracy_alpha}
    \vspace{-.2in}
  \end{minipage}
  \hfill
  \begin{minipage}[b]{0.23\textwidth}
    \centering
    \includegraphics[width=\textwidth]{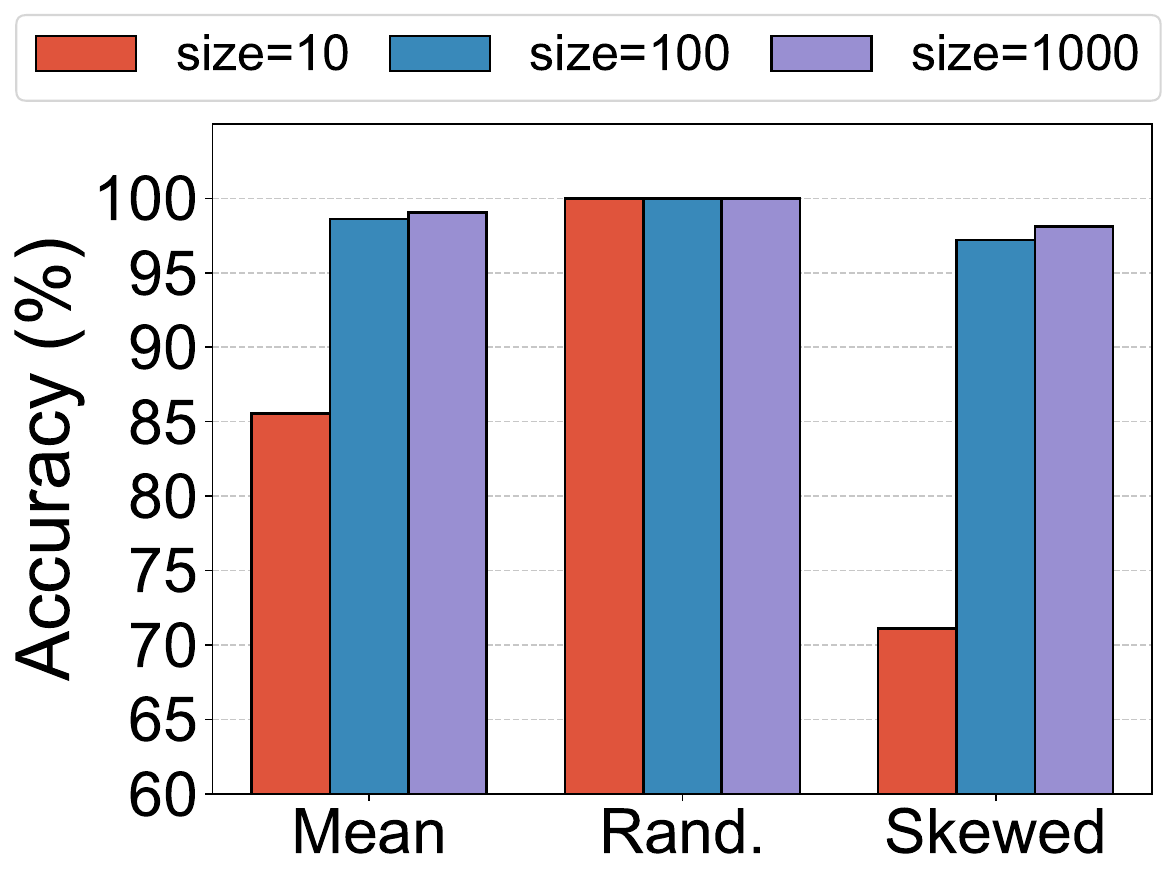}
    \captionof{figure}{Impact of observation window size.}
    \vspace{-.2in}
    \label{accuracy_size}
  \end{minipage}
\end{figure}




\phm{Impact of Cache Size.} We discuss the performance of \athena \ under different cache size configurations. Apart from the cache size, all other experimental settings are kept consistent with those used in the end-to-end evaluation (\Cref{sec:eval_macro}). 
As shown in \Cref{chr_with_sizes}, \athena \ outperforms JuiceFS across all the evaluated cache size settings.
In general, the smaller the cache size, the higher the performance benefit of \athena. 
In particular, even when the cache size is sufficient (100\%), our method still remarkably outperform JuiceFS (96.2\% versus 64.3\% in CHR); this is because by automatic prefetching, we can substantially reduce the compulsory cache misses.

\phm{Overhead.}We also measure the additional computation and memory overhead brought by \athena.
As shown in \Cref{overhead}, we vary the number of nodes in \texttt{AccessStreamTree} to different magnitudes, and measure the performance as well as time and memory overheads. 
With more nodes, the pattern recognition can be made more accurate and the overall JCT would be smaller.
Yet, more nodes in \texttt{AccessStreamTree} in the meantime incur a larger overhead---for computation overhead we can observe a logarithmic increase and for memory overhead the trend is linear---both consistent with our complexity analysis in \Cref{sec:implementation}.
In particular, at the default setup (10000 nodes), the additional computation overhead amortized to each I/O request is 47.6 µs (only 0.36\% of the average I/O time of 13.2 ms), and the memory cost to maintain \texttt{AccessStreamTree} is 73.2 MB, also acceptable given typical cache configuration and the performance benefit of \athena. 

\begin{figure}
  \centering
  \begin{minipage}[b]{0.23\textwidth}
    \centering
    \includegraphics[width=\textwidth]{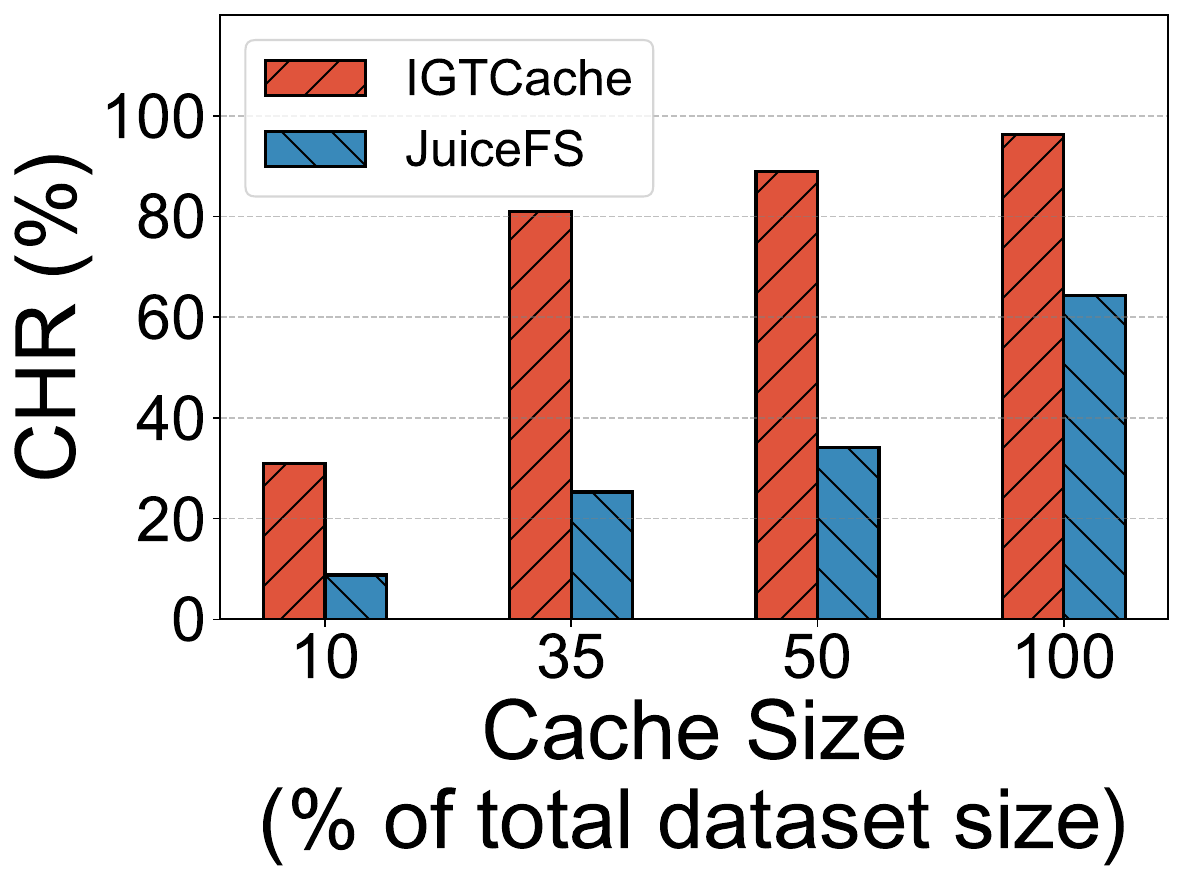}
    \vspace{-0.1in}
    \captionof{figure}{CHR under different cache sizes.}
    \label{chr_with_sizes}
    \vspace{-.2in}
  \end{minipage}
  \hfill
  \begin{minipage}[b]{0.23\textwidth}
      \vspace{-0.1in}
    \centering
    \includegraphics[width=\textwidth]{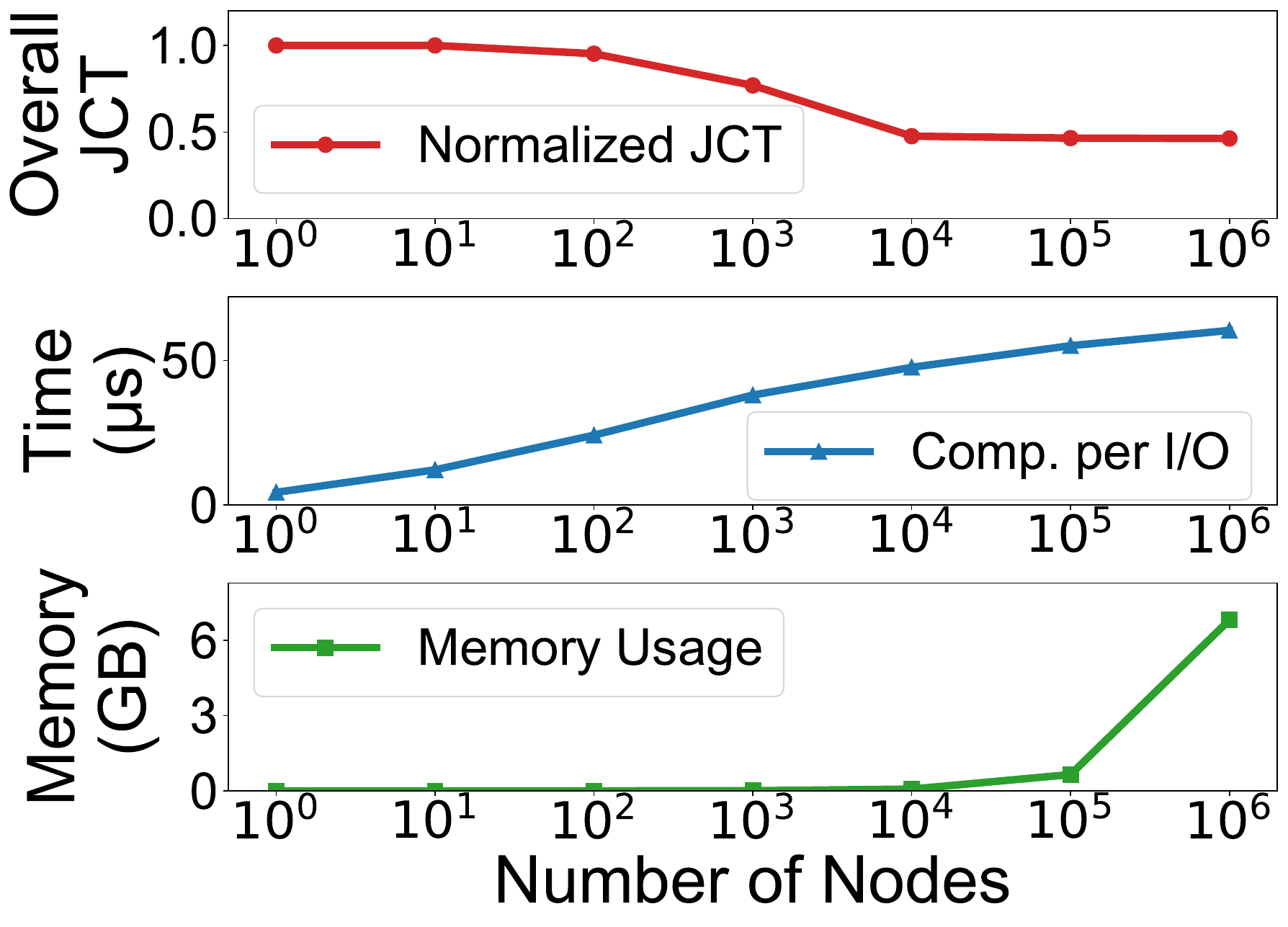}
    \captionof{figure}{Overhead with different node numbers.}
    \label{overhead}
    \vspace{-.2in}
  \end{minipage}
\end{figure}

\section{Conclusion}
\label{sec:conclusion}

In this paper, we present \athena, a unified, high-efficacy cache for modern AI clusters.
\athena\ adapts to the diverse data storage granularities and data access patterns in a user-transparent manner.
Specifically, it models the data requests using an \texttt{AccessStreamTree} and categorize the access sequence of each \texttt{AccessStream} to a specific pattern. 
Based on that pattern, \athena\ customizes the caching strategies for prefetching, eviction, and allocation to attain high caching efficacy.
Testbed experiments across a mixture of diverse workloads show that, \athena\ effectively improves the caching efficiency over existing approaches with negligible overhead.  

\bibliographystyle{myplain}
\bibliography{main}

\end{document}